\documentclass[10pt]{article}
\usepackage[preprint]{tmlr}

\usepackage{amsmath,amsfonts,bm}

\def\eqref#1{equation~\ref{#1}}

\def\1{\bm{1}}

\DeclareMathAlphabet{\mathsfit}{\encodingdefault}{\sfdefault}{m}{sl}
\SetMathAlphabet{\mathsfit}{bold}{\encodingdefault}{\sfdefault}{bx}{n}

\usepackage[T1]{fontenc}
\usepackage{hyperref}
\hypersetup{colorlinks=true, linkcolor=blue, urlcolor=blue, citecolor=blue}
\usepackage{url}
\usepackage{xspace}
\usepackage{tabularx}
\usepackage{multirow}
\usepackage{graphicx}
\usepackage{booktabs}
\usepackage{enumitem}
\usepackage{amsmath}
\usepackage{amsfonts}
\usepackage{amssymb}
\usepackage{txfonts}
\usepackage{bigstrut,rotating}
\usepackage{colortbl}
\usepackage{pifont}
\usepackage{algorithm}
\usepackage{algpseudocode}
\usepackage{calc}
\usepackage{array}
\usepackage{makecell}
\usepackage{subcaption}
\usepackage{float}
\usepackage{adjustbox}
\usepackage{fontawesome5}

\usepackage{xcolor}
\usepackage{tikz}
\usetikzlibrary{svg.path}
\usepackage{listings}
\usepackage{courier}
\usepackage{tcolorbox}
\tcbuselibrary{skins,breakable}

\definecolor{titleblue}{RGB}{0,0,160}
\definecolor{bglightblue}{RGB}{240,240,255}
\definecolor{codebg}{RGB}{232,232,255}
\definecolor{hfYellow}{HTML}{FFD21E}
\definecolor{hfOrange}{HTML}{FF9D0B}
\definecolor{hfDark}{HTML}{3A3B45}
\definecolor{hfRed}{HTML}{FF323D}

\newtcolorbox{expbox}[1]{%
  enhanced,
  colback=bglightblue,
  colframe=titleblue,
  boxrule=1.5pt,
  arc=10pt,
    left=10pt,
    right=10pt,
    top=18pt,
    bottom=8pt,
  boxsep=0pt,
    drop fuzzy shadow=black!18,
  before skip=15pt,
  after skip=15pt,
  fontupper=\normalfont,
    attach boxed title to top left={yshift=-1.5mm, xshift=8mm},
  boxed title style={
    colback=titleblue,
    colframe=titleblue,
    arc=5pt,
    boxrule=0pt,
    left=10pt,
    right=10pt,
    top=6pt,
    bottom=6pt,
    fontupper=\bfseries\large\color{white}
  },
  title={#1}
}

\newtcolorbox{hashoutputbox}{%
    enhanced,
    width=0.99\linewidth,
    box align=center,
    colback=white,
    colframe=titleblue,
    boxrule=1pt,
    arc=4pt,
    left=10pt,
    right=10pt,
    top=8pt,
    bottom=8pt,
    boxsep=0pt,
    before skip=4pt,
    after skip=0pt
}

\newcommand{\stepitem}[1]{\noindent\textbf{Step #1:}~}

\newcommand{\codetext}[1]{\texttt{#1}}

\title{H3D: Benchmarking Unsupervised Text Hashing for Fine-Grained Document Deduplication}

\author{\normalfont
\begin{tabular*}{\textwidth}{@{}l@{\extracolsep{\fill}}r@{}}
\textbf{Qianren Mao}$^{*}$ & \textit{maogr@zgclab.edu.cn} \\
\textit{Zhongguancun Laboratory} & \\[0.4em]
\textbf{Jiaxun Lyu} & \textit{19373403@buaa.edu.cn} \\
\textit{Beihang University} & \\[0.4em]
\textbf{Junnan Liu} & \textit{junnan.liu@monash.edu} \\
\textit{Monash University Australia} & \\[0.4em]
\textbf{Zhijun Chen} & \textit{zhijun.chen@polyu.edu.hk} \\
\textit{Hong Kong Polytechnic University} & \\[0.4em]
\textbf{Jingzheng Li} & \textit{ljz@zgclab.edu.cn} \\
\textit{Zhongguancun Laboratory} & \\[0.4em]
\textbf{Hanwen Hao} & \textit{20373190@buaa.edu.cn} \\
\textit{Beihang University} & \\[0.4em]
\textbf{Bo Li} & \textit{libo@act.buaa.edu.cn} \\
\textit{Beihang University} & \\[1.0em]
\multicolumn{2}{@{}l@{}}{\small $^{*}$Corresponding author.}
\end{tabular*}}

\def\month{07}
\def\year{2026}
\def\openreview{\url{https://openreview.net}}

\begin{document}
\maketitle
\begin{abstract}
Document hashing provides compact representations for efficient similarity search and document deduplication, but existing studies rarely compare hashing pipelines under a unified protocol for fine-grained scientific documents. H3D is an unsupervised text hashing benchmark for fine-grained document deduplication. It evaluates representative unsupervised non-learning hashing approaches (MinHash, SimHash, Winnowing, FuzzyHash, FlyHash) together with semantic-sensitive methods built from frozen BGE embeddings and two quantization strategies (BGE-BIHash and BGE-LSHash). The non-learning methods generate hash fingerprints through manually designed mathematical rules without training or labeled similarity pairs, which distinguishes them from neural semantic hashing models. We benchmark all methods on CSFCube and RELISH, two datasets that provide complementary evaluation settings: facet-level analysis for scientific-document similarity and larger-scale split-level evaluation for biomedical similarity search. H3D jointly reports ranking quality (MAP, NDCG@20), efficiency, and robustness under controlled text compression. The results show a consistent trade-off: lexical and structural fingerprints are competitive for near-duplicate matching, while semantic-sensitive representations better preserve similarity under content rewriting, at higher computational cost. We further analyze when different similarity measures become rank-equivalent for specific hash representations, improving the interpretability and reproducibility of method comparisons.

\end{abstract}
\begin{center}
{\normalsize
\href{https://github.com/DocAILab/Document-Fingerprints}{\raisebox{-0.22em}{\includegraphics[height=1.32em]{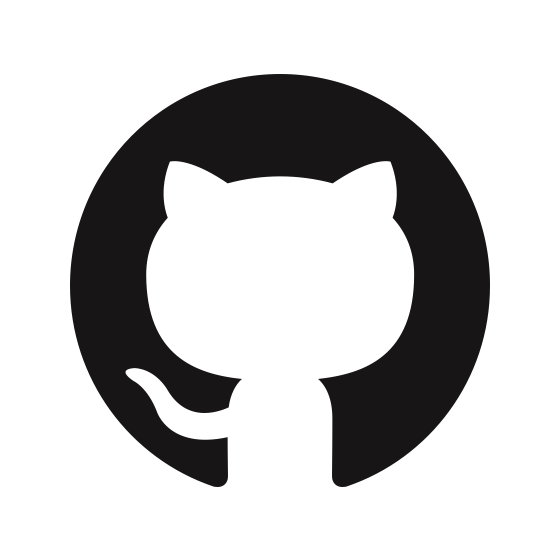}}\ \textbf{\texttt{DocAILab/Document-Fingerprints}}}
\quad\quad
\href{https://huggingface.co/datasets/DocAILab/Document-Fingerprints}{\raisebox{-0.20em}{\includegraphics[height=1.18em]{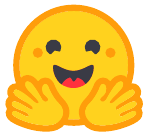}}\ \textbf{\texttt{Hugging Face Dataset}}}
}
\end{center}

\section{Introduction}
Document hashing is a fundamental technique for efficient textual similarity search~\cite{HuangP0YP24} and document deduplication~\cite{LeeINZECC22}. 
It enables the transformation of high-dimensional text representations into compact hash codes while preserving semantic or structural properties, allowing for fast and scalable document comparison. 
At web and pretraining-corpus scale, a representative non-learning pipeline is MinHashLSH: documents are first converted into token or n-gram sets, MinHash signatures approximate their Jaccard overlap, and locality-sensitive hashing splits the signature into bands so that highly overlapping documents are likely to collide in at least one bucket~\cite{RaoZ16,AliyunMinHashLSH}. This design turns exhaustive pairwise comparison into candidate generation followed by lightweight verification, which explains its continued use in large-scale text-cleaning systems. Recent systems further optimize this pipeline for extreme-scale deduplication. For example, LSHBloom replaces the conventional MinHashLSH index with Bloom filters, preserving MinHashLSH-style duplicate detection while substantially reducing storage requirements for billion-document settings~\cite{KhanUSBAHGBCHF24}. These developments show that practical deduplication depends not only on the hash signature itself, but also on the indexing and candidate-generation mechanism wrapped around it.
In recent years, learning-based hashing methods~\cite{ChaidaroonF17,HenaoCSSWWC18,DongSSC19,ZhengSSC20,DoanR20,QiuSOYC21,OuSYZZL21,QiuSYS22}, including supervised and unsupervised neural variants, have improved retrieval accuracy in many reported settings. 
Supervised document deduplication techniques have notable drawbacks~\cite{NunesHVAVNG23}: they require costly labeled datasets and often exhibit domain specificity, limiting their adaptability. Practical deduplication tasks also involve nuanced similarities for which accurate labels are difficult to obtain. These limitations motivate hashing methods that function without labeled data, increasing interest in unsupervised approaches.

Unsupervised document hashing eliminates the need for labeled similarity pairs and is therefore appealing for large-scale applications where labeled training data is scarce or unavailable~\cite{abs-2102-08942}. 
Within this broad setting, it is useful to distinguish three families. First, \emph{unsupervised non-learning hashing} methods use fixed mathematical rules and handcrafted transformations to produce fingerprints, such as MinHash~\cite{RaoZ16,ChristianiP17}, SimHash~\cite{JiangS11}, Winnowing~\cite{SchleimerWA03}, FuzzyHash, and FlyHash. These methods require no training corpus, no labeled similarity pairs, and no learned parameters, which makes them reproducible and efficient baselines for near-duplicate detection. Second, \emph{deep-learning unsupervised semantic hashing} methods learn discrete or quantized document representations from unlabeled corpora. Representative examples include VDSH, which uses variational inference to learn latent binary semantic codes~\cite{ChaidaroonF17}; NASH, which introduces an end-to-end neural architecture for generative semantic hashing~\cite{HenaoCSSWWC18}; BMSH and CorrSH, which refine generative hashing with mixture priors and Boltzmann-machine-enhanced dependencies~\cite{DongSSC19,ZhengSSC20}; DABA, which uses an adversarial autoencoder for implicit unsupervised text hashing~\cite{DoanR20}; BOTTLENECK, which learns hash codes through a contrastive information bottleneck~\cite{QiuSOYC21}; DHIM, which refines BERT embeddings by maximizing mutual information~\cite{OuSYZZL21}; and MICPQ, which performs end-to-end refinement and contrastive product quantization over BERT embeddings~\cite{QiuSYS22}. AEPQ, CSH, and related contrastive quantization variants follow the same direction: their encoders or quantizers are optimized with reconstruction, adversarial, mutual-information, or contrastive objectives. Third, recent embedding-based pipelines use pretrained text encoders and apply lightweight binarization or locality-sensitive hashing on top of frozen embeddings.

In this work, H3D focuses on the first and third families: non-learning hash baselines and BGE-based frozen-embedding quantization. Deep-learning unsupervised hashing models are not benchmarked as training-based competitors because they introduce additional variables that are outside the scope of this protocol, including training corpus selection, reconstruction or contrastive objectives, optimization budget, architecture choice, and model-selection criteria. Excluding them allows H3D to isolate the behavior of hash generators, similarity functions, and frozen pretrained embeddings under the same query--candidate evaluation pipeline. Semantic-agnostic non-learning hashing methods are efficient for detecting exact or near-exact duplicates, but they can struggle with more nuanced cases, such as paraphrased content or semantically equivalent expressions. 
Semantic-sensitive hashing methods based on frozen pretrained embeddings preserve deeper semantic relationships before compact quantization. 
Despite these advancements, the field lacks a unified benchmark for evaluating text hashing techniques in fine-grained document deduplication. Existing studies primarily focus on general document similarity or retrieval tasks, leaving a gap in benchmarking hashing methods for identifying subtle textual variations. A standardized evaluation framework is needed to compare different hashing techniques and analyze their strengths and weaknesses in fine-grained deduplication scenarios.

To address these challenges, we make the following key contributions:

\begin{itemize}[leftmargin=2em]
    \item We introduce \textbf{H3D}, an unsupervised text \textbf{h}ashing benchmark for fine-grained \textbf{d}ocument \textbf{d}eduplication \textbf{d}etection, covering CSFCube and RELISH under a unified query--candidate protocol.
    
    \item We conduct a controlled empirical study of unsupervised non-learning hashing methods across string-, set-, and vector-based similarity families, characterizing method--metric compatibility within a shared evaluation pipeline.
    
    \item We provide a unified quantization view for BGE-based hashing, instantiate it with BGE-BIHash and BGE-LSHash, and analyze accuracy, efficiency, and robustness under the same benchmark pipeline.
\end{itemize}

Together, these contributions position H3D as a diagnostic benchmark rather than a single-method proposal: the goal is to clarify when a hashing family is useful, which scorer should be paired with it, and what accuracy--efficiency trade-off is introduced by compact semantic quantization.

\begin{figure*}[t]
    \centering
    \includegraphics[width=0.70\textwidth]{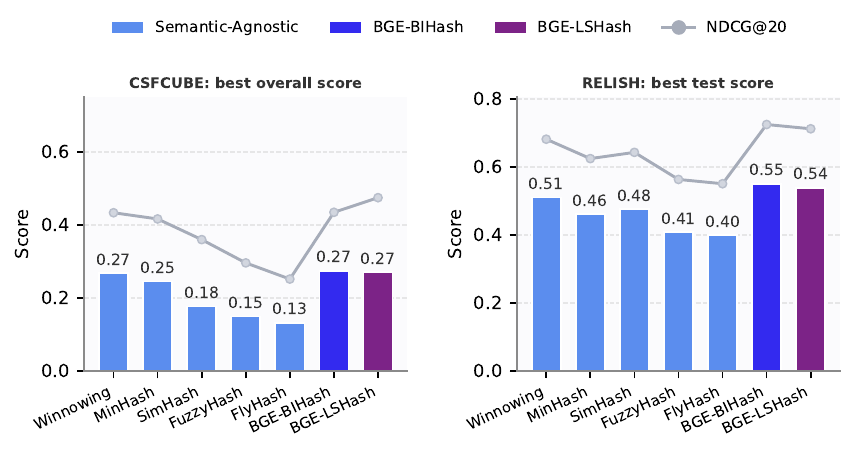}
    \caption{
        Overview of best-performing hashing configurations on CSFCUBE and RELISH, generated from the experimental summaries and BGE result tables. Bars report MAP and lines report NDCG@20. Light-blue bars denote unsupervised non-learning hashing methods, while blue and purple bars denote BGE-BIHash and BGE-LSHash.
    }
    \label{fig:h3d_benchmark_overview}
\end{figure*}

Figure~\ref{fig:h3d_benchmark_overview} provides a compact overview of the benchmark outcomes by comparing representative best-performing configurations across CSFCUBE and RELISH. The figure is intended as a visual entry point to the later result sections, where the same trends are analyzed in detail by dataset, facet, hashing family, and similarity function.

We focus on the following research questions:
{\small
\begin{itemize}[leftmargin=2em]
    \item \textbf{RQ1}: How much does metric choice change semantic-agnostic hash ranking quality?
    \item \textbf{RQ2}: How much retrieval quality remains after quantizing dense BGE embeddings?
    \item \textbf{RQ3}: Under progressive text compression, which hashing paradigm is more robust, and at what computational cost?
\end{itemize}
}

\section{Benchmarking D3 Datasets}
H3D uses two public scientific-document similarity datasets: CSFCube~\cite{CSFCUBE} and RELISH~\cite{RELISH}. Both datasets provide titles, abstracts, document identifiers, and graded relevance annotations, but they emphasize different evaluation settings. CSFCube is designed for faceted query-by-example search over computer science papers, whereas RELISH focuses on expert-curated similarity detection in biomedical literature. This complementarity allows H3D to evaluate whether a hashing method preserves fine-grained facet-level similarity and whether it remains effective in a larger-scale query--candidate ranking setting.

\subsection{Dataset Statistics}

\begin{table}[t]
\scriptsize
\caption{Dataset statistics used in the H3D query--candidate ranking benchmark.}
\renewcommand\arraystretch{1.2}
\setlength{\abovecaptionskip}{0.5mm}
\resizebox{\linewidth}{!}{
\begin{tabular}{lcccccccc}
\toprule
\textbf{Dataset} & \textbf{Lang.} & \textbf{Corpus} & \textbf{Queries} & \makecell{\textbf{Query abstract-}\\\textbf{facet pairs}} & \makecell{\textbf{Query-candidate}\\\textbf{pairs}} & \textbf{Avg. words} & \textbf{Labels} & \textbf{Facets} \\ 
\midrule
CSFCube  & English & 4,207 & 50 & 50 & 6,244 & 167.3 & 0--3 & background/method/result/overall \\
RELISH   & English & 163,170 & 3,274 & 3,190 & 191,245 & 241.3 & 0--2 & overall \\
\bottomrule
\end{tabular}}
\label{tab:dataset_statistic}
\end{table}

The basic statistics of these datasets are shown in Table~\ref{tab:dataset_statistic}.
The corpus and annotation statistics are computed from the official CSFCube release\footnote{\url{https://github.com/iesl/CSFCube}} and the RELISH-Aspire release\footnote{\url{https://figshare.com/articles/dataset/RELISH-Aspire/19425506}}. Since H3D is formulated as query--candidate ranking rather than supervised model training, Table~\ref{tab:dataset_statistic} reports query abstract-facet units and query-candidate annotation pairs instead of train/dev/test counts.

\textbf{CSFCube.} CSFCube is a test collection for faceted query-by-example retrieval over computer science research papers~\cite{CSFCUBE}. Its official release contains a corpus of paper titles and abstracts, a set of query papers, and facet-specific relevance annotations. The core design of CSFCube is that a query paper may be similar to a candidate paper for different reasons: it may share the same research background, use a similar method, or report related results. Accordingly, the benchmark provides graded labels for background, method, and result facets, together with an overall view used in our experiments. The release also provides predicted sentence-level discourse labels for abstracts, including background, objective, method, result, and other labels. We use these labels to organize abstract content into facet-aware text fields before applying hash generation and similarity scoring.

\textbf{RELISH.} RELISH is an expert-curated biomedical literature similarity benchmark introduced by Brown and Zhou~\cite{RELISH}. The dataset was designed to evaluate document similarity detection in biomedical search, where related papers may differ in wording while still describing closely related biomedical findings, methods, or study contexts. The released RELISH-Aspire data contains paper titles and abstracts, query metadata, candidate pools, and graded relevance annotations. Unlike CSFCube, RELISH does not provide explicit facet-level labels; each query is evaluated as an overall document similarity task. This makes RELISH useful for testing whether hashing methods scale from small, facet-rich scientific corpora to a larger biomedical ranking setting.

For both datasets, we extract titles and abstracts as the main textual fields for fingerprint generation. For CSFCube, background/objective, method, and result-related sentences are additionally grouped according to the provided predicted labels. For RELISH, incomplete records are filtered before evaluation, yielding 3,190 annotated query units and 191,245 query-candidate pairs in our processed benchmark.

\subsection{Dataset Analysis}

Although both CSFCube and RELISH are useful resources for benchmarking document hashing techniques, they stress different aspects of fine-grained document deduplication.

CSFCube provides a structured evaluation framework by explicitly categorizing similarity into background, method, and result facets. This structured labeling allows for an in-depth analysis of document similarity along multiple dimensions, making it particularly useful for assessing semantic-sensitive hashing techniques that rely on contextual embeddings. However, the dataset's relatively small scale and manual labeling constraints limit its applicability for large-scale hashing evaluations.

RELISH presents a larger and more realistic similarity landscape, covering varied textual transformations, citation-based relatedness, and document revisions. The dataset reflects practical deduplication challenges such as distinguishing preprint and final-publication variants, handling paraphrased but semantically close research, and identifying different versions of the same paper with minor edits. These properties make RELISH well suited for evaluating both semantic-agnostic and semantic-sensitive hashing methods at scale. Unlike CSFCube, RELISH lacks explicit facet-based similarity labels, which makes it less suitable for isolating specific similarity dimensions.

Together, CSFCube and RELISH provide complementary evaluation settings. CSFCube supports detailed fine-grained similarity analysis through explicit facet-level annotations, whereas RELISH provides a larger testbed that reflects real-world document duplication scenarios. Their combination supports evaluation across different levels of textual variation and duplication.

\section{Benchmarking Text Hash Methods}

Text hashing supports efficient document retrieval and deduplication by replacing full text with compact fingerprints or codes. The usefulness of a hash depends on what information it preserves: some methods preserve lexical overlap and local structure, while others preserve contextual semantics before compression. H3D therefore benchmarks two primary categories of text hashing techniques: semantic-agnostic non-learning hashing, which focuses on structural and lexical properties through manually specified rules, and semantic-sensitive hashing, which leverages pretrained language models for deeper semantic representation.

\subsection{Benchmark Scope}
The benchmark covers methods implemented and evaluated in the same experimental pipeline. H3D includes representative unsupervised non-learning hashing approaches (\textit{MinHash}, \textit{SimHash}, \textit{Winnowing}, \textit{FuzzyHash}, and \textit{FlyHash}) and BGE-based frozen-embedding hashing variants (\textit{BGE-BIHash} and \textit{BGE-LSHash}). The first group generates fingerprints through manually designed mathematical rules, while the second group first obtains pretrained BGE embeddings and then applies deterministic binarization or locality-sensitive hashing. Neither group requires task-specific training on the benchmark datasets.

\begin{table*}[t]
    \centering
    \small
    \renewcommand\arraystretch{1.2}
    \setlength{\tabcolsep}{2.4mm}
    \caption{Benchmark-scope distinction among unsupervised hashing families. H3D evaluates fixed-rule hashes and frozen-embedding hash pipelines, but excludes models that require task-specific neural training.}
    \label{tab:hashing_scope}
    \resizebox{\textwidth}{!}{
    \begin{tabular}{llllc}
        \toprule
        \textbf{Hashing category} & \textbf{Network training} & \textbf{Pretrained encoder} & \textbf{Hash module} & \textbf{Included} \\
        \midrule
        Non-learning hashing & None & Not used & Fixed handcrafted rules & \ding{51} \\
        Deep unsupervised hashing & Required & Optional or fine-tuned & Learnable neural hash layer & \ding{55} \\
        Frozen embedding + LSH hashing & None & Frozen, not fine-tuned & Post-hoc binarization or LSH & \ding{51} \\
        \bottomrule
    \end{tabular}}
\end{table*}

Table~\ref{tab:hashing_scope} distinguishes benchmark scope rather than performance. The non-learning group includes methods such as MinHash, SimHash, Winnowing, FuzzyHash, and FlyHash; they mainly target efficient near-duplicate detection and lexical or structural overlap matching. The frozen-embedding group includes BGE-BIHash, BGE-LSHash, and SemHash-LLM-style pipelines; these methods reuse pretrained semantic embeddings and apply a lightweight hash step without updating the encoder. In contrast, deep-learning unsupervised hashing methods such as VDSH, NASH, BMSH, CorrSH, DABA, BOTTLENECK, DHIM, and MICPQ represent a progression from variational and generative semantic hashing toward BERT-based mutual-information and contrastive quantization. Their common feature is that the hash representation is learned from data through an optimization objective.

This distinction defines the benchmark scope. Deep-learning unsupervised hashing models are important related work, but their main emphasis is usually semantic document retrieval or compact binary indexing, where the central question is whether a learned latent code improves retrieval quality after training. They are therefore not directly plug-and-play for H3D's fine-grained deduplication protocol: adapting them would require selecting or constructing an unlabeled training corpus, deciding whether to fine-tune BERT-like encoders, choosing a code length and optimization objective, and retraining before evaluation. Including them would turn H3D into a model-training benchmark. In contrast, the methods covered by H3D can be applied directly to the same query--candidate pools without task-specific optimization, so the comparison centers on hash construction, scorer compatibility, runtime, and robustness.

\subsection{Semantic-Agnostic Non-Learning Hash}
Representative unsupervised non-learning hashing approaches include \textit{FlyHash}, \textit{MinHash}, \textit{SimHash}, \textit{Winnowing}, and \textit{FuzzyHash}. They serve as baselines for fine-grained document deduplication because they generate hash fingerprints through manually designed mathematical rules without training or labeled similarity pairs. This distinguishes them from deep unsupervised semantic hashing models trained with self-supervised or contrastive objectives.

\begin{table*}[h]
    \centering
    \small
    \renewcommand\arraystretch{1.2}
    \setlength{\tabcolsep}{1.5mm}
    \caption{Summary of Semantic-Agnostic Non-Learning Hash Methods}
    \label{tab:hash_methods}
    \resizebox{\textwidth}{!}{
    \begin{tabular}{lcccccc}
        \toprule
        \textbf{Method} & \textbf{Non-linearity} & \textbf{Bit Balance} & \textbf{Bit Independence} & \textbf{Data Independence} & \textbf{Dimensional Consistency} & \textbf{Position Invariance} \\ 
        \midrule
        Minhash     & \ding{55} & -- & \ding{55} & \ding{55} & \ding{51} & \ding{51} \\
        Karp-Rabin  & \ding{55} & \ding{55} & \ding{55} & \ding{51} & \ding{55} & \ding{55} \\
        Winnowing   & \ding{55} & \ding{55} & \ding{51} & \ding{51} & \ding{55} & \ding{55} \\
        Simhash     & \ding{55} & \ding{51} & \ding{55} & \ding{55} & \ding{51} & \ding{51} \\
        LSH         & \ding{55} & \ding{51} & \ding{55} & \ding{55} & \ding{51} & \ding{51} \\
        Fuzzyhash   & \ding{55} & \ding{55} & \ding{51} & \ding{51} & \ding{55} & \ding{55} \\
        Flyhash     & \ding{51} & \ding{55} & \ding{51} & \ding{55} & \ding{51} & \ding{55} \\
        \bottomrule
    \end{tabular}}
\end{table*}

\subsubsection{Fundamental Principles and Features}
\textbf{Minhash}: Proposed by Broder~\cite{Minhash}, Minhash estimates the Jaccard similarity of sets by applying multiple hash functions and recording the minimum values. It is efficient for processing large-scale set-based data and was originally used in the AltaVista search engine for duplicate detection. In practical MinHashLSH pipelines, text is converted into n-gram sets, MinHash produces a compact signature, and LSH banding maps signatures into buckets so that only bucket-sharing documents become candidate duplicates~\cite{RaoZ16,AliyunMinHashLSH}. This makes MinHashLSH attractive for large corpus cleaning, while newer variants such as LSHBloom target the memory cost of the LSH index by replacing it with Bloom-filter-based candidate generation~\cite{KhanUSBAHGBCHF24}. However, Minhash remains tied to lexical set overlap and is less effective when long documents share meaning but diverge substantially in wording or organization. 
\textbf{Karp-Rabin}: Introduced by Rabin and Karp~\cite{Karp-Rabin} in 1987, this algorithm employs a sliding window approach to compute hash values for substrings and compares them to a pattern's hash value. While effective for exact substring matching, its sensitivity to text order and inability to handle flexible similarity make it less applicable to fingerprinting tasks.
\textbf{Winnowing}: Proposed by Schleimer et al.~\cite{Winnowing}, Winnowing enhances the Karp-Rabin algorithm by selecting the minimum hash values within sliding windows, mitigating sensitivity to text order and location. It generates robust fingerprints for large-scale documents but struggles with short texts due to insufficient feature capture.
\textbf{Simhash}: Developed by Manku et al. ~\cite{Simhash}, Simhash employs feature extraction by tokenizing text, assigning weights, and applying binary hashing to generate fixed-length fingerprints. While effective for texts with over 500 tokens, its inter-bit dependency may reduce discrimination in certain applications.
\textbf{Locality Sensitive Hashing} (LSH): Introduced by Indyk et al.~\cite{LSH}, LSH maps high-dimensional data to hash buckets based on similarity. By converting comparisons into bucket matches, it efficiently handles large-scale and high-dimensional data. Its versatility allows integration with Minhash and Simhash for improved performance.
\textbf{Fuzzy Hashing}: Proposed by Kornblum~\cite{Fuzzyhash}, Fuzzy Hashing combines weak and strong hash functions to segment text and compute segment hashes. While robust to text order changes, it is computationally expensive and unsuitable for large datasets. The resulting fingerprint is a length-proportional ASCII string.
\textbf{FlyHash}: Inspired by the fruit fly olfactory system, FlyHash was proposed by Dasgupta et al.~\cite{Flyhash}. It uses high-dimensional binary mapping and a Winner-Take-All mechanism to produce sparse binary vectors. Despite its computational efficiency, it requires careful preprocessing to ensure vector consistency in fingerprinting tasks.

\subsubsection{Analysis of Semantic-Agnostic Non-Learning Hashing}
Table~\ref{tab:hash_methods} summarizes the key properties of these semantic-agnostic non-learning hashing methods.
The practical behavior of these non-learning hashes is representation-dependent. Methods such as MinHash and Winnowing primarily encode lexical overlap patterns, while SimHash and FlyHash encode projected token-feature signals in binary spaces. Consequently, retrieval quality depends jointly on the hash generator and the scorer applied to its outputs.

Empirically, these methods remain competitive for near-duplicate matching where lexical structure is largely preserved. However, because their fingerprints are rule-based rather than learned from document neighborhoods, they are less robust when semantically similar documents diverge substantially in wording or organization. This motivates semantic-sensitive hashing, where dense encoders provide a semantic prior before quantization.

\subsection{Semantic-Sensitive Hash}

The non-learning hash baselines above primarily rely on statistical or structural properties of text, which limits their ability to capture deep semantic relationships. In contrast, semantic-sensitive hashing first maps documents into a representation space where paraphrases and semantically related descriptions can remain close even when surface forms differ. Pretrained language models (PLMs) provide such contextual vector spaces, but directly using dense PLM embeddings for document deduplication presents several challenges, including high dimensionality, computational cost, and the absence of a discrete fingerprint. In H3D, the semantic-sensitive category therefore refers to BGE-based hashing methods, where dense BGE embeddings are converted into compact hash representations through post-hoc quantization.

\subsubsection{Adapting BGE for Document Hashing}

The BAAI General Embedding (BGE) model is a Transformer-based language model trained for retrieval, producing dense vectors that capture fine-grained semantics. Unlike semantic-agnostic hashes that rely on lexical statistics, BGE maps documents into a continuous semantic space. Dense embeddings are expensive to store and compare at scale, so H3D studies compact binary quantization on top of BGE.

We use a unified two-stage formulation:
\begin{equation}
e(x) = f_\theta(x) \in \mathbb{R}^D, \qquad b(x) = Q(e(x)) \in \{0,1\}^L,
\end{equation}
where $f_\theta$ is a frozen BGE encoder, $D$ is the embedding dimension, $Q$ is a quantizer, and $L$ is the target binary-code length. Retrieval is performed by a scorer $s(b(q), b(d))$.

The first approach, \textit{BGE-BIHash}, applies deterministic pooling and sign-based binarization. The normalized embedding is reshaped into an $m\times L$ matrix with $m=D/L$, so that the implementation can aggregate the $D$ coordinates into $L$ groups and then threshold them into an $L$-bit code. This reshape step is an implementation convenience rather than a new semantic transform: it preserves the original embedding values while aligning them with the downstream pooling layout. In our main BGE-base configuration, we use the common $D=1024$ embedding dimension and set $L=32$, which gives $m=32$ and therefore the concrete $32\times32$ layout. Algorithm~\ref{alg:bge_bihash} instantiates this BIHash quantizer.

The second approach, \textit{BGE-LSHash}, applies random-hyperplane LSH to the same normalized BGE embeddings. It partitions the embedding into $k$ equal subspaces of dimension $D/k$, applies $r=L/k$ random projections per subspace, and concatenates the resulting bits into an $L$-bit signature. Compared with BIHash, this stochastic quantizer is designed to preserve neighborhood structure in expectation, but introduces additional hyperparameters (e.g., number of subspaces and projections) and runtime overhead. Algorithm~\ref{alg:bge_lsh} instantiates this LSHash quantizer.

BGE-BIHash and BGE-LSHash share the same semantic source but differ in quantization bias: BIHash favors deterministic compression, while LSHash favors neighborhood-sensitive random projection. Our experiments compare these choices under identical evaluation settings.



\newcommand{\CommentRight}[1]{\hfill{\footnotesize\texttt{// #1}}}

\begin{algorithm}[t]
\footnotesize
\caption{Document Fingerprint Generation based on BGE-BIHash}
\label{alg:bge_bihash}
\begin{algorithmic}[1]
\Procedure{DocumentFingerprint}{$text, D, L$}
    \State $m \gets D / L$ \CommentRight{number of coordinates pooled per bit}
    \State $\mathbf{v} \gets \textsc{BGEEncoder}(text, D)$ \CommentRight{obtain normalized BGE embedding}
    \State $\mathbf{V} \gets \operatorname{reshape}(\mathbf{v}, m, L)$ \CommentRight{reshape $\mathbf{v}$ into an $m \times L$ matrix}
    \State $\mathbf{a} \gets (1/m)\sum_{i=1}^{m}\mathbf{V}_{i,:}$ \CommentRight{compute column-wise averages}
    \State $\mathbf{f} \gets \operatorname{array}[L]$ \CommentRight{initialize binary fingerprint}
    \For{$j \in \{1, \dotsc, L\}$}
        \State $f[j] \gets
        \begin{cases}
        1, & \text{if } a[j] > 0,\\
        0, & \text{otherwise}
        \end{cases}$ \CommentRight{threshold-based binarization}
    \EndFor
    \State \Return $\mathbf{f}$ \CommentRight{output $L$-bit fingerprint}
\EndProcedure

\vspace{0.4em}
\Procedure{BGEEncoder}{$text, D$}
    \State $\mathbf{u} \in \mathbb{R}^{D} \gets \textsc{BGEEmbedding}(text)$ \CommentRight{obtain BGE embedding}
    \State $\mathbf{v} \gets \mathbf{u} / \|\mathbf{u}\|_2$ \CommentRight{L2 normalization}
    \State \Return $\mathbf{v}$
\EndProcedure
\end{algorithmic}
\end{algorithm}

\begin{algorithm}[t]
\footnotesize
\caption{Document Fingerprint Generation based on BGE-LSHash}
\label{alg:bge_lsh}
\begin{algorithmic}[1]
\Procedure{DocumentFingerprint}{$\text{texts}, D, k, L$}
    \State $r \gets L / k$ \CommentRight{number of projections per subvector}
    \State Sample $r$ random hyperplanes $\{\mathbf{h}_1, \dotsc, \mathbf{h}_r\} \subset \mathbb{R}^{D/k}$ \CommentRight{random LSH directions}
    \State $\mathbf{fingerprints} \gets [\,]$ \CommentRight{initialize fingerprint list}
    \For{$t \in \text{texts}$}
        \State $\mathbf{v} \gets \textsc{BGEEncoder}(t, D)$ \CommentRight{obtain normalized BGE embedding}
        \State Partition $\mathbf{v}$ into $\{\mathbf{v}_1, \dotsc, \mathbf{v}_k\}$ \CommentRight{divide into $k$ equal subvectors}
        \State $\mathbf{b} \gets \text{LSHEncode}(\{\mathbf{v}_i\}, \{\mathbf{h}_j\})$ \CommentRight{hash each subvector}
        \State Append $\mathbf{b}$ to $\mathbf{fingerprints}$ \CommentRight{store document fingerprint}
    \EndFor
    \State \Return $\mathbf{fingerprints}$ \CommentRight{output binary fingerprints}
\EndProcedure

\vspace{0.4em}
\Procedure{LSHEncode}{$\{\mathbf{v}_i\}, \{\mathbf{h}_j\}$}
    \State Initialize $\mathbf{b} \gets$ "" \CommentRight{empty binary code}
    \For{$i = 1$ to $k$}
        \For{$j = 1$ to $r$}
            \State $h_{i,j} \gets \mathbb{I}[\langle \mathbf{v}_i, \mathbf{h}_j \rangle \geq 0]$ \CommentRight{1 if projection $\geq$ 0, else 0}
        \EndFor
        \State Concatenate $\{h_{i1}, \dotsc, h_{ir}\}$ to $\mathbf{b}$ \CommentRight{form subvector code}
    \EndFor
    \State \Return $\mathbf{b}$ \CommentRight{final fingerprint string}
\EndProcedure
\end{algorithmic}
\end{algorithm}

\subsection{Implications for Fine-Grained Document Deduplication}

Semantic-sensitive hashing methods can improve robustness to paraphrasing by leveraging contextual representations. However, the choice of quantization strategy directly affects their applicability to fine-grained document deduplication. BGE-BIHash prioritizes storage efficiency and deterministic encoding, making it suitable for scenarios where compact fingerprints are required. In contrast, BGE-LSHash is designed to better preserve semantic proximity within the hash space, which can benefit similarity retrieval when local neighborhoods in the embedding space are informative.

BGE-based quantization connects semantic-rich representations with scalable document deduplication methods. The experiments below compare these variants with non-learning hash baselines under the same ranking protocol.

\section{Benchmarking Evaluation Protocol}

The evaluation protocol defines a common query--candidate ranking task, shared metrics, and type-specific similarity functions for heterogeneous hash outputs. This design ensures that methods with different output forms can be compared through the same retrieval interface.

\subsection{Problem Modeling}
Each hashing method is evaluated as a document similarity model. The key question is how well the hash representation preserves information needed for ranking while reducing the dimensionality or storage cost of the original text representation.

Let $\mathcal{Q}$ denote the query set. For each query $q \in \mathcal{Q}$, let $\mathcal{D}_q = \{d_1,\dots,d_{N_q}\}$ be its candidate documents with relevance labels $R_q(d)$. A hashing method defines an encoder
\begin{equation}
H: x \mapsto \mathbf{Z} \in \mathcal{Z},
\end{equation}
where $\mathcal{Z}$ may be a binary string, set/multiset, or vector space. A scorer then produces
\begin{equation}
S_q(d) = S\big(H(q), H(d)\big),
\end{equation}
and candidates are ranked in descending order of $S_q(d)$.
Equation (3) defines the common representation interface, and Equation (4) defines the scalar score used for ranking under that interface.

To assess whether a hash preserves information useful for similarity matching, we evaluate the ranking quality of the retrieved documents. For each query, the benchmark obtains a ranked list of candidate documents based on computed similarity scores and compares that list with the dataset's ground-truth relevance labels. Ranking quality is quantified with Mean Average Precision (MAP) and Normalized Discounted Cumulative Gain (NDCG), which capture both retrieval accuracy and the placement of relevant documents near the top of the list.

\subsection{Evaluation Metrics}
To quantitatively assess retrieval quality, we report Mean Average Precision (MAP) and NDCG@20 under a unified query--candidate ranking protocol. For each query $q\in\mathcal{Q}$, the scorer outputs a ranked candidate list
with $\Pi_q=(d_{q,1}, d_{q,2},\dots,d_{q,N_q})$ and $N_q=|\mathcal{D}_q|$, where $d_{q,i}$ is the document at rank $i$ and graded relevance is $R_{q,i}=y_q(d_{q,i})$. For binary-relevance metrics, we use $\mathrm{REL}_q(i)=\mathbb{1}[R_{q,i}\ge g_0]$ with threshold $g_0=2$ for both CSFCube and RELISH.

The core AP/MAP definition is
\begin{equation}
\mathrm{AP}(q)=\big(\max(1,\sum_{i=1}^{N_q}\mathrm{REL}_q(i))\big)^{-1}\sum_{k=1}^{N_q} P_q(k)\,\mathrm{REL}_q(k),
\end{equation}
where $P_q(k)=(1/k)\sum_{i=1}^{k}\mathrm{REL}_q(i)$ and $\mathrm{MAP}=(1/|\mathcal{Q}|)\sum_{q\in\mathcal{Q}}\mathrm{AP}(q)$.

For graded relevance, we use NDCG@20:
\begin{equation}
\mathrm{NDCG}@20=(1/|\mathcal{Q}|)\sum_{q\in\mathcal{Q}}\big(\mathrm{DCG}_q@20/\mathrm{IDCG}_q@20\big),
\end{equation}
with linear-gain definitions $\mathrm{DCG}_q@20=\sum_{i=1}^{20}\big(R_{q,i}/\log_2(i+1)\big)$ and $\mathrm{IDCG}_q@20=\sum_{i=1}^{\min(20,N_q)}\big(R^*_{q,i}/\log_2(i+1)\big)$, where $R^*_{q,i}$ is sorted by descending relevance.

\textbf{Implementation note.} The evaluation script first writes query-level metrics and then reports split/facet-level means (test/dev for RELISH; facet-wise for CSFCube). Besides MAP and NDCG@20, the code also computes NDCG@K, precision@K, recall@K, and F1@K; we keep MAP and NDCG@20 as the main comparable indicators because they are available for all reported configurations and summarize complementary ranking properties.

\subsection{Similarity Methods}
Different hashing methods transform documents into different output types, so similarity functions are selected according to the resulting representation. We use three groups of similarity measures: string-based, set-based, and vector-based measures. Each group corresponds to specific hash-output properties, as summarized in Table~\ref{tab:hashing_similarity}.

For a fixed query--candidate pair $(q,d)$, we compute hash outputs $Z_q=H(q)$ and $Z_d=H(d)$. Because $Z$ may be a string, set/multiset, or vector, we use a type-specific adapter $\Pi_\tau(\cdot)$ and scorer $S_\tau(\cdot,\cdot)$ under the unified rule
\begin{equation}
S(q,d) = S_\tau\!\left(\Pi_\tau(Z_q),\,\Pi_\tau(Z_d)\right).
\end{equation}
Here, $\tau\in\{\text{str},\text{set},\text{mset},\text{vec}\}$ indicates representation type.
The adapter $\Pi_\tau$ is a deterministic conversion step that maps raw hash output to the input domain expected by a similarity function:
\begin{itemize}[leftmargin=2em]
\item $\Pi_{\text{str}}$: keep or serialize fingerprints as strings (e.g., fixed-length bit strings, variable-length ASCII fingerprints).
\item $\Pi_{\text{set}}$: convert list-like outputs to sets for duplicate-insensitive overlap.
\item $\Pi_{\text{mset}}$: convert outputs to multiplicity counters $c(t)$ for duplicate-aware overlap.
\item $\Pi_{\text{vec}}$: convert outputs to numeric vectors of aligned dimension (including padding/truncation when needed).
\end{itemize}
In the implementation, this adapter behavior is realized inside the unified similarity interface (string/list-to-vector conversion and type dispatch), so all methods can be ranked under one common pipeline.

\subsubsection{String-Based Similarity Methods}
String-based similarity methods operate directly on string representations; they are most suitable for hashing techniques that produce binary or ASCII strings. The two most common approaches are:
\begin{itemize}[leftmargin=2em]
    \item Hamming Distance: for equal-length hashes with $\tilde{Z}_q=\Pi_{\text{str}}(Z_q)$ and $\tilde{Z}_d=\Pi_{\text{str}}(Z_d)$, we use $d_H(\tilde{Z}_q,\tilde{Z}_d)=\sum_{i=1}^{n}\mathbb{1}(\tilde{Z}_q[i]\neq\tilde{Z}_d[i])$ and convert distance to score by $S_H(q,d)=1/(1+d_H(\tilde{Z}_q,\tilde{Z}_d))$.

    \item Levenshtein Distance: for variable-length strings, edit distance $d_L(\tilde{Z}_q,\tilde{Z}_d)$ counts minimum insertions, deletions, and substitutions; the ranking score is $S_L(q,d)=1/(1+d_L(\tilde{Z}_q,\tilde{Z}_d))$.
    This is used for outputs such as Winnowing and FuzzyHash fingerprints.

\end{itemize}

\subsubsection{Set-Based Similarity Methods}
Set-based similarity measures operate on unordered collections of hashed features, making them ideal for hashing techniques that produce sets rather than sequences. Two key measures in this category are Jaccard similarity and Multiset Jaccard similarity.

\begin{itemize}[leftmargin=2em]
    \item \textbf{Jaccard Similarity}: with $A_q=\Pi_{\text{set}}(Z_q)$ and $A_d=\Pi_{\text{set}}(Z_d)$, we use $S_J(q,d)=J(A_q,A_d)=|A_q\cap A_d|/|A_q\cup A_d|$.
    This measure is particularly effective for \textit{MinHash}, as MinHash is designed to preserve set overlap. In code, list-like signatures are cast to sets before computing overlap.

    \item \textbf{Multiset Jaccard}: to preserve duplicate counts, we use $S_{J_m}(q,d)=\big(\sum_t\min(c_q(t),c_d(t))\big)/\big(\sum_t\max(c_q(t),c_d(t))\big)$, where $c_q(t)$ and $c_d(t)$ are multiplicities after multiset adaptation $\Pi_{\text{mset}}$.
\end{itemize}

\subsubsection{Vector-Based Similarity Methods}
Vector-based similarity measures operate on continuous or high-dimensional binary vectors. We convert distances to ranking scores so that larger values always indicate higher similarity.

\begin{itemize}[leftmargin=2em]
    \item \textbf{Cosine Similarity}: with $\mathbf{V}_q=\Pi_{\text{vec}}(Z_q)$ and $\mathbf{V}_d=\Pi_{\text{vec}}(Z_d)$, we use the standard cosine score $S_{\cos}(q,d)=(\mathbf{V}_q^\top \mathbf{V}_d)/(\|\mathbf{V}_q\|_2\,\|\mathbf{V}_d\|_2)$.
    \item \textbf{Manhattan Score}: with $d_1(\mathbf{V}_q,\mathbf{V}_d)=\|\mathbf{V}_q-\mathbf{V}_d\|_1$, we define $S_{L1}(q,d)=1/(1+d_1(\mathbf{V}_q,\mathbf{V}_d))$.
    \item \textbf{Mahalanobis Score}: with covariance $\Sigma$, we define $S_M(q,d)=\exp\!\left(-0.5\,(\mathbf{V}_q-\mathbf{V}_d)^\top\Sigma^{-1}(\mathbf{V}_q-\mathbf{V}_d)\right)$.
    In our default implementation for general vectors, $\Sigma=I$; for binary vectors, a diagonal covariance proxy is used. In both cases, higher $S_M$ means more similar pairs.
\end{itemize}

\begin{table*}
    \centering
    \small
    \renewcommand\arraystretch{1.2}
    \setlength{\tabcolsep}{1.2mm}
    \caption{Comparison of Hashing Methods and Suitable Similarity Measures}
    \label{tab:hashing_similarity}
    \resizebox{\textwidth}{!}{
    \begin{tabular}{lccc}
        \toprule
        \textbf{Hashing Method} & \textbf{Output Type} & \textbf{Recommended Similarity Measures} & \textbf{Key Characteristics} \\ 
        \midrule
        SimHash     & Fixed-length binary string  & Hamming Distance & Bitwise comparison, sensitive to minor changes \\
        Winnowing   & Variable-length string  & Edit Distance  & Localized substring hashing, length variation \\
        Winnowing (Converted) & Multiset  & Multiset Jaccard Similarity & Position-sensitive, benefits from set-based comparison \\
        MinHash     & Set of hashed shingles  & Jaccard Similarity  & Preserves Jaccard similarity, suitable for set-based data \\
        FlyHash     & Sparse binary vector  & Cosine Similarity  & High-dimensional encoding, angle-based comparison \\
        BGE Hashing & Dense semantic embedding  & Manhattan Score, Cosine Similarity & Continuous vector representation, preserves semantics \\
        \bottomrule
    \end{tabular}}
\end{table*}

\subsection{Experimental Setup and Details}
All experiments were conducted in the same Python evaluation pipeline with method-specific dependencies. We treat the BGE embedding dimension $D$ and hash length $L$ as method parameters; in the main BGE-base setting used for the BIHash layout, we choose the common $D=1024$ dense vector and report fingerprints at 32, 64, and 128 bits depending on the method. For each query, the benchmark scores every candidate in the pool and ranks candidates with the selected scorer. The reported \textit{Total Time} values come from the scoring scripts in the same codebase and should be interpreted as relative cost within this benchmark rather than hardware-agnostic throughput.

The benchmark code fixes the split policy and several method-specific parameters. CSFCube uses the full query set as test data, while RELISH loads test/dev membership from \textit{relish-evaluation\_splits.json} and aggregates metrics separately. LSH-BGE uses seed 123 in hash-projection generation, fc-BGE uses \texttt{np.random.seed(123)} for random projection initialization, and FlyHash uses seed 55 in sparse random projection. The main defaults include hash dimension 128, n-gram size 3, and Winnowing window 5. BGE backbones are selected from \{small, base, large\}, while LSH-BGE exposes the number of tables $k$ and hash size/bit budget $L$ as command-line arguments; the heatmap analysis script evaluates a grid over multiple $k$ and $L$ combinations. For BGE-based methods, \texttt{FlagModel} automatically selects \texttt{cuda}, \texttt{mps}, \texttt{npu}, or \texttt{cpu} according to availability, and fp16 is enabled only when supported.

\section{Results and Analyses}

The experiments compare unsupervised non-learning hash methods and semantic-sensitive BGE-based hashing methods, including BGE-BIHash and BGE-LSHash. All methods are evaluated on \textbf{RELISH} and \textbf{CSFCUBE}, two scientific-document datasets with similarity labels. The analysis is organized around the three research questions: scorer sensitivity for non-learning hashes, quality retention after BGE quantization, and robustness under controlled text compression.

\subsection{Semantic-Agnostic Non-Learning Hash Analysis (RQ1)}
The first set of experiments evaluates representative unsupervised non-learning hashing methods, including \textit{SimHash}, \textit{MinHash}, \textit{Winnowing}, \textit{FuzzyHash}, and \textit{FlyHash}. The evaluation reports \textbf{Mean Average Precision (MAP)}, \textbf{Normalized Discounted Cumulative Gain (NDCG@20)}, and execution time. Because these methods produce different output types, the central issue is not only which hash generator is strongest, but also which similarity function is compatible with each representation.

Table~\ref{tab:csfcube_results} summarizes the results on the CSFCUBE dataset.\footnote{Some similarity measures become rank-equivalent only for specific hash representations. For binary signatures such as FlyHash and SimHash, Hamming distance, Manhattan distance, and Mahalanobis distance (under identity covariance) are monotonic transforms of mismatch counts, so they produce the same ranking and therefore identical MAP/NDCG; we keep Manhattan as the representative row. For MinHash in this implementation, Hamming/Jaccard/Levenshtein also collapse to the same ranking, so we keep Jaccard as the representative row. This behavior does not appear uniformly in other hashing methods because their outputs are not constrained to the same binary/count structure (e.g., weighted multiset fingerprints or real-valued vectors), so different similarity functions emphasize different aspects and can change document ordering.}
Because CSFCube provides facet-level annotations, it allows us to inspect whether a hashing method behaves consistently across background, method, result, and overall similarity. The results show clear variation across facets: background similarity generally receives higher scores, whereas method-level similarity is more difficult for most semantic-agnostic methods. These results suggest that surface-oriented fingerprints remain useful for near-duplicate scientific texts, but their effectiveness depends strongly on the selected similarity measure and the evaluated facet.


\begin{table*}[!htbp]
\centering
\small
\renewcommand\arraystretch{1.2}
\setlength{\tabcolsep}{1.2mm}
\caption{Performance comparison of document hashing methods on the \textbf{CSFCUBE} dataset.}
\label{tab:csfcube_results}
\resizebox{\textwidth}{!}{
\begin{tabular}{@{}llcccccccccr@{}}
\toprule
\textbf{Hashing Method} & \textbf{Similarity Measure} &
\multicolumn{2}{c}{\textbf{Background}} &
\multicolumn{2}{c}{\textbf{Method}} &
\multicolumn{2}{c}{\textbf{Result}} &
\multicolumn{2}{c}{\textbf{All}} &
\textbf{Total Time (s)} \\
\cmidrule(lr){3-4} \cmidrule(lr){5-6} \cmidrule(lr){7-8} \cmidrule(lr){9-10}
& & MAP & NDCG & MAP & NDCG & MAP & NDCG & MAP & NDCG &  \\
\midrule
\textbf{FlyHash} & cosine & 0.1690 & 0.2853 & 0.0872 & 0.1976 & 0.1148 & 0.2256 & 0.1228 & 0.2352 & 13.66 \\
& jaccard & 0.1827 & 0.3483 & 0.1006 & 0.1965 & 0.1172 & 0.2167 & 0.1325 & 0.2520 & 10.95 \\
& levenshtein & 0.1591 & 0.2893 & 0.1031 & 0.2270 & 0.1327 & 0.2167 & 0.1311 & 0.2434 & 10.57 \\
& manhattan & 0.1692 & 0.2826 & 0.0852 & 0.1969 & 0.1151 & 0.2291 & 0.1222 & 0.2353 & 11.67 \\
& M-jaccard & 0.1682 & 0.2929 & 0.0993 & 0.2055 & 0.1090 & 0.2113 & 0.1246 & 0.2355 & 11.34 \\
\midrule
\textbf{FuzzyHash} & cosine & 0.1931 & 0.3516 & 0.0978 & 0.2188 & 0.1478 & 0.2670 & 0.1453 & 0.2777 & 0.28 \\
& jaccard & 0.1931 & 0.3516 & 0.0978 & 0.2188 & 0.1478 & 0.2670 & 0.1453 & 0.2777 & 0.28 \\
& levenshtein & 0.2213 & 0.3988 & 0.0896 & 0.2211 & 0.1425 & 0.2750 & 0.1497 & 0.2963 & 0.25 \\
& mahalanobis & 0.1892 & 0.3604 & 0.1006 & 0.1965 & 0.1251 & 0.2249 & 0.1373 & 0.2586 & 15.33 \\
& manhattan & 0.1850 & 0.3230 & 0.1054 & 0.2229 & 0.1271 & 0.2518 & 0.1382 & 0.2647 & 0.27 \\
& M-jaccard & 0.1879 & 0.3486 & 0.0977 & 0.2315 & 0.1515 & 0.2530 & 0.1449 & 0.2763 & 0.52 \\
\midrule
\textbf{MinHash} & cosine & 0.2320 & 0.4098 & 0.1240 & 0.2369 & 0.1908 & 0.3365 & 0.1813 & 0.3261 & 18.98 \\
& jaccard & 0.3469 & 0.5383 & 0.1678 & 0.3216 & 0.2266 & 0.3974 & 0.2451 & 0.4167 & 17.31 \\
& mahalanobis & 0.1892 & 0.3604 & 0.1006 & 0.1965 & 0.1251 & 0.2249 & 0.1373 & 0.2586 & 87.87 \\
& manhattan & 0.2169 & 0.3877 & 0.1013 & 0.2272 & 0.1423 & 0.2511 & 0.1523 & 0.2867 & 18.91 \\
\midrule
\textbf{SimHash} & cosine & 0.2446 & 0.4545 & 0.1177 & 0.2711 & 0.1643 & 0.3458 & 0.1741 & 0.3552 & 5.16 \\
& manhattan & 0.2464 & 0.4592 & 0.1189 & 0.2769 & 0.1697 & 0.3500 & 0.1770 & 0.3601 & 3.62 \\
& levenshtein & 0.2048 & 0.3773 & 0.1357 & 0.2585 & 0.1613 & 0.2834 & 0.1665 & 0.3050 & 3.19 \\
& M-jaccard & 0.1885 & 0.3476 & 0.0960 & 0.1890 & 0.1443 & 0.2358 & 0.1420 & 0.2557 & 3.63 \\
\midrule
\textbf{Winnowing} & cosine & 0.2006 & 0.3497 & 0.0952 & 0.2227 & 0.1330 & 0.2394 & 0.1418 & 0.2690 & 5.22 \\
& jaccard & 0.3820 & 0.5741 & 0.1659 & 0.3137 & 0.2620 & 0.4216 & 0.2678 & 0.4337 & 6.22 \\
& levenshtein & 0.3277 & 0.5211 & 0.1517 & 0.2971 & 0.2459 & 0.4169 & 0.2401 & 0.4095 & 2.89 \\
& mahalanobis & 0.1892 & 0.3604 & 0.1006 & 0.1965 & 0.1251 & 0.2249 & 0.1373 & 0.2586 & 47.74 \\
& manhattan & 0.1944 & 0.3267 & 0.0903 & 0.1997 & 0.1246 & 0.2294 & 0.1353 & 0.2505 & 5.89 \\
& M-jaccard & 0.3695 & 0.5811 & 0.1499 & 0.3126 & 0.2443 & 0.4181 & 0.2522 & 0.4344 & 7.42 \\
\bottomrule
\end{tabular}}
\end{table*}

\begin{figure*}[!htbp]
    \centering
    \includegraphics[width=0.92\textwidth]{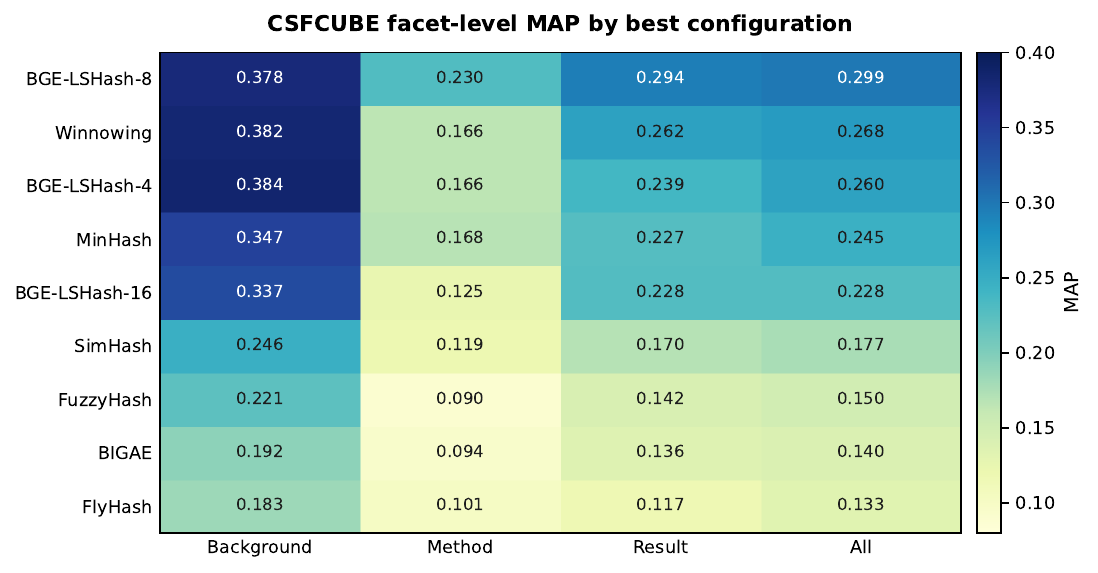}
    \caption{
        Facet-level MAP comparison on CSFCUBE. Each row reports the best configuration for a hashing method, and columns correspond to background, method, result, and overall similarity facets.
    }
    \label{fig:csfcube_facet_map_heatmap}
\end{figure*}

Figure~\ref{fig:csfcube_facet_map_heatmap} further visualizes the best configuration of each method across the CSFCUBE facets.
The background facet generally receives higher MAP scores than the method and result facets, indicating that broad topical similarity is easier to preserve than fine-grained methodological similarity.
The heatmap also shows that the best-performing method can change across facets, so a single overall score may hide important differences in fine-grained similarity behavior.

\begin{figure*}[!htbp]
    \centering
    \includegraphics[width=0.92\textwidth]{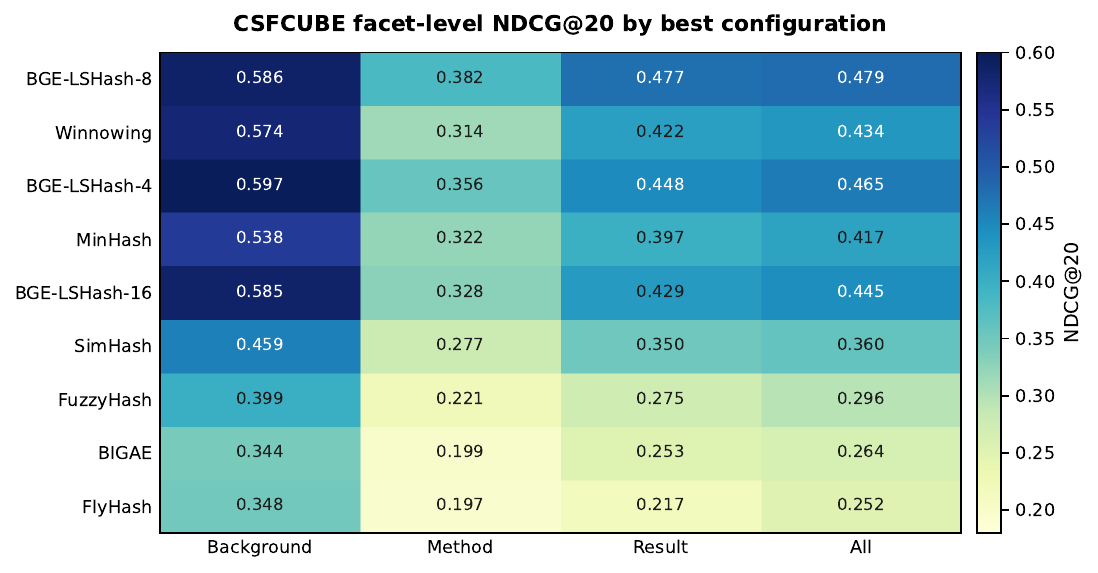}
    \caption{
        Facet-level NDCG@20 comparison on CSFCUBE. Each row reports the best configuration for a hashing method, and columns correspond to background, method, result, and overall similarity facets.
    }
    \label{fig:csfcube_facet_ndcg_heatmap}
\end{figure*}

Figure~\ref{fig:csfcube_facet_ndcg_heatmap} provides the corresponding NDCG@20 view.
The trend is consistent with MAP: background similarity is easier to rank correctly, while method-level similarity remains the most challenging facet.
The NDCG@20 view complements MAP by emphasizing the quality of the top-ranked candidates.

\begin{figure*}[!htbp]
    \centering
    \includegraphics[width=0.92\textwidth]{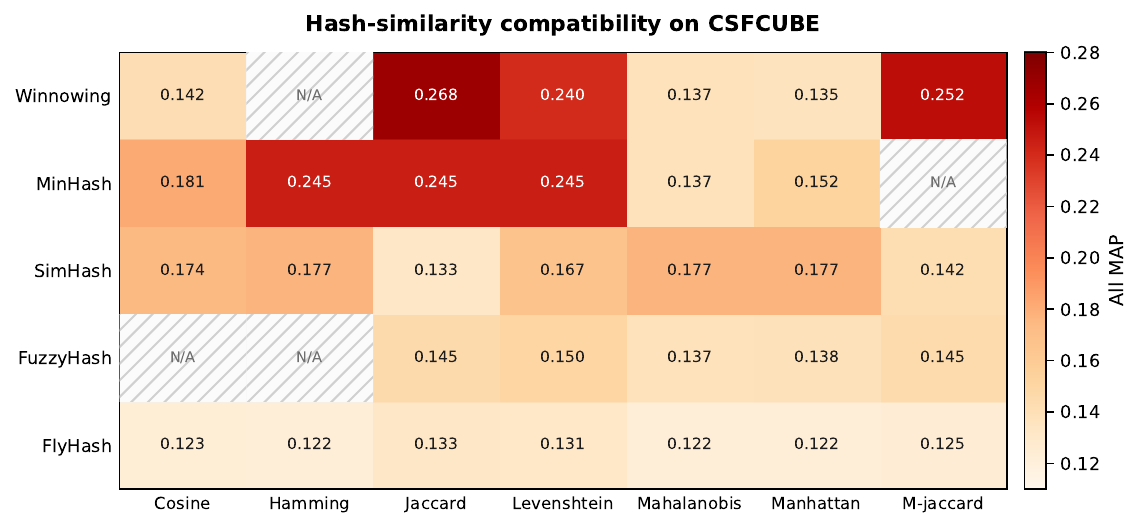}
    \caption{
        Compatibility between hash methods and similarity measures on CSFCUBE. Values denote overall MAP, and hatched N/A cells indicate unavailable or invalid method--metric combinations in the current experimental outputs.
    }
    \label{fig:hash_similarity_map_heatmap}
\end{figure*}

Figure~\ref{fig:hash_similarity_map_heatmap} summarizes the compatibility between semantic-agnostic hash representations and similarity measures.
The results show that the best similarity metric is method-dependent, with no single similarity measure dominating all hash representations.
These results motivate joint evaluation of hash generation and similarity measurement rather than separate treatment of the two components.

\begin{figure*}[!htbp]
    \centering
    \includegraphics[width=0.92\textwidth]{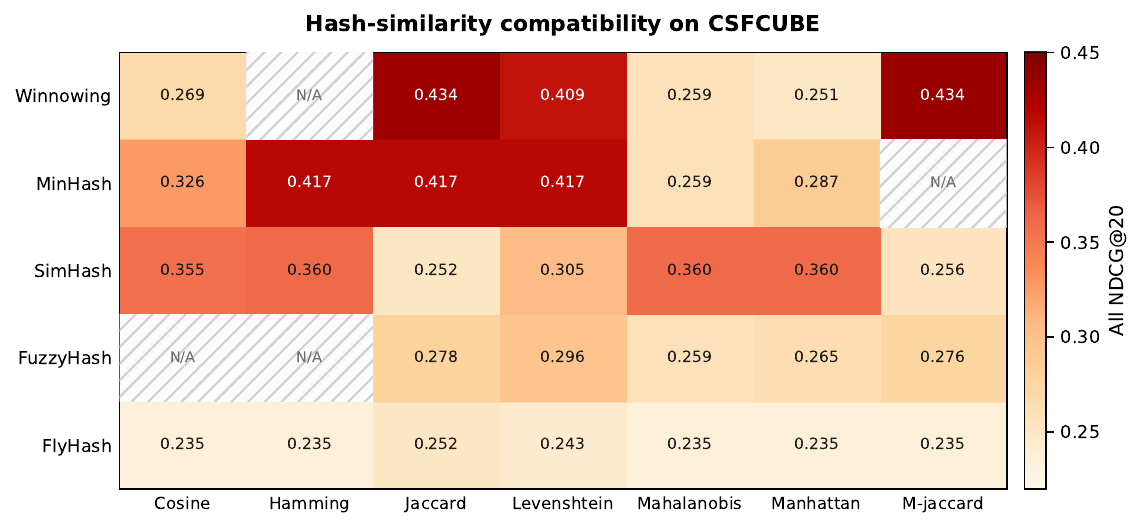}
    \caption{
        Compatibility between hash methods and similarity measures on CSFCUBE. Values denote overall NDCG@20, and hatched N/A cells indicate unavailable or invalid method--metric combinations in the current experimental outputs.
    }
    \label{fig:hash_similarity_ndcg_heatmap}
\end{figure*}

Figure~\ref{fig:hash_similarity_ndcg_heatmap} provides the corresponding NDCG@20 compatibility view.
The pattern is broadly consistent with the MAP heatmap, indicating that compatibility between a hash representation and a similarity measure is empirical rather than purely definitional.
In practice, the choice of similarity measure affects both relevance coverage and top-rank ordering quality, so reporting a hash method without its scorer can be misleading.

For Table~\ref{tab:relish_results}, we apply the same deduplication rule as Table~\ref{tab:csfcube_results}: when different similarity measures produce identical ranking outcomes for a given hash representation, we retain one representative row only.

\begin{table*}[!htbp]
\centering
\small
\renewcommand\arraystretch{1.2}
\setlength{\tabcolsep}{1.2mm}
\caption{Performance comparison of document hashing methods with different similarity measures on the \textbf{RELISH} dataset (combined results).}
\label{tab:relish_results}
\begin{tabular}{llcccccc}
\toprule
\textbf{Hashing Method} & \textbf{Similarity Measure} &
\multicolumn{2}{c}{\textbf{Test}} &
\multicolumn{2}{c}{\textbf{Dev}} &
\textbf{Total Time (s)} \\
\cmidrule(lr){3-4} \cmidrule(lr){5-6}
& & MAP & NDCG@20 & MAP & NDCG@20 & \\
\midrule

\textbf{FlyHash} & cosine & 0.4002 & 0.5506 & 0.4036 & 0.5613 & 0.00 \\
& hamming & 0.4171 & 0.5736 & 0.4149 & 0.5757 & 1738.03 \\
& levenshtein & 0.4305 & 0.5969 & 0.4306 & 0.6025 & 1744.65 \\
\midrule
\textbf{FuzzyHash} & cosine & 0.3994 & 0.5516 & 0.4008 & 0.5576 & 84.66 \\
& jaccard & 0.4064 & 0.5611 & 0.4044 & 0.5642 & 14.14 \\
& levenshtein & 0.4091 & 0.5632 & 0.4119 & 0.5721 & 13.91 \\
& M-jaccard & 0.4070 & 0.5607 & 0.4022 & 0.5613 & 23.88 \\
\midrule

\textbf{MinHash} & cosine & 0.4101 & 0.5643 & 0.4130 & 0.5748 & 0.00 \\
& jaccard & 0.4609 & 0.6245 & 0.4614 & 0.6310 & 4159.82 \\
& mahalanobis & 0.9747 & 1.0000 & 0.9726 & 1.0000 & 4426.86 \\
& manhattan & 0.4080 & 0.5614 & 0.4104 & 0.5679 & 4077.91 \\
\midrule
\textbf{SimHash} & cosine & 0.4578 & 0.6200 & 0.4596 & 0.6258 & 0.00 \\
& hamming & 0.3771 & 0.5168 & 0.3805 & 0.5294 & 1254.07 \\
& levenshtein & 0.4026 & 0.5579 & 0.4017 & 0.5660 & 1190.81 \\
& manhattan & 0.4756 & 0.6428 & 0.4768 & 0.6487 & 0.00 \\
& M-jaccard & 0.4498 & 0.6232 & 0.4513 & 0.6337 & 1232.29 \\
\midrule
\textbf{Winnowing} & jaccard & 0.4876 & 0.6551 & 0.4944 & 0.6601 & 1339.80 \\
& levenshtein & 0.3693 & 0.4915 & 0.3728 & 0.5031 & 1328.98 \\
& M-jaccard & 0.4996 & 0.6694 & 0.5011 & 0.6758 & 1454.27 \\
\bottomrule
\end{tabular}
\end{table*}

\begin{figure*}[!htbp]
    \centering
    \includegraphics[width=0.72\textwidth]{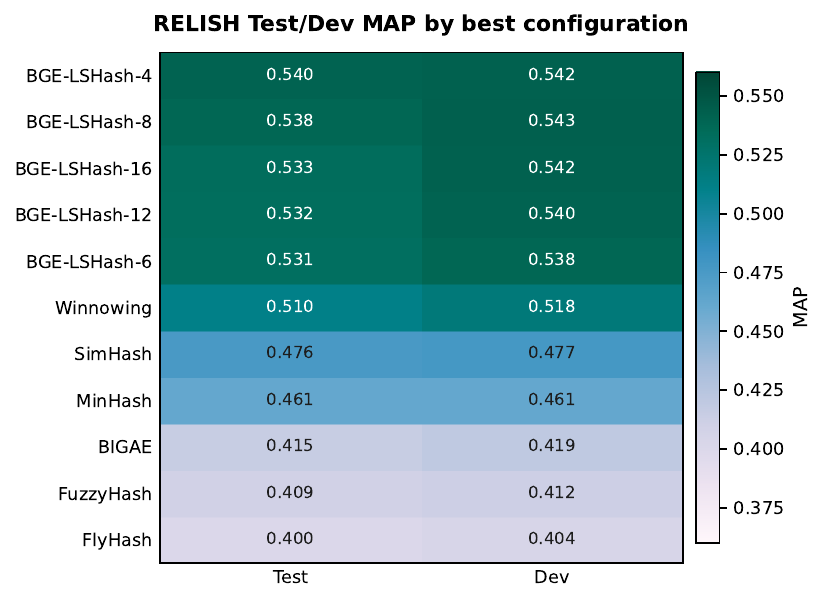}
    \caption{
        RELISH Test/Dev MAP comparison. Each row reports the best configuration for a hashing method according to Test MAP.
    }
    \label{fig:relish_split_map_heatmap}
\end{figure*}

\begin{figure*}[!htbp]
    \centering
    \includegraphics[width=0.72\textwidth]{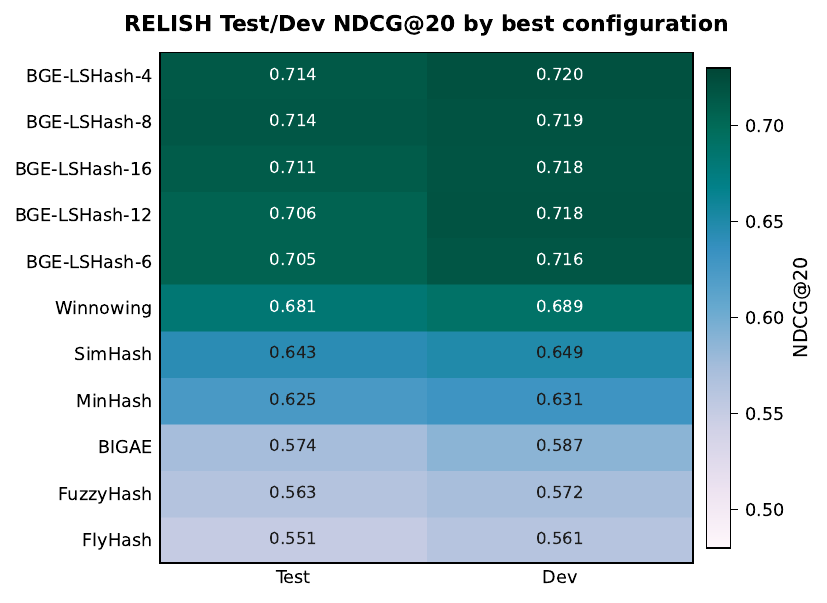}
    \caption{
        RELISH Test/Dev NDCG@20 comparison. The close Test/Dev values indicate that the ranking behavior is largely consistent across splits.
    }
    \label{fig:relish_split_ndcg_heatmap}
\end{figure*}

Figure~\ref{fig:relish_split_map_heatmap} and Figure~\ref{fig:relish_split_ndcg_heatmap} summarize the RELISH results across the test and development splits.
Unlike CSFCUBE, RELISH does not provide facet-level labels, so the visualization focuses on split-level consistency.
The two heatmaps show that the relative ranking patterns are broadly similar between the test and development splits, which suggests that the observed method differences are not driven by a single split artifact.

\subsection{Semantic-Sensitive Hash Analysis (RQ2)}
We evaluate semantic-sensitive hashing methods using the \textbf{BAAI General Embedding (BGE)} family~\cite{bge_embedding}. Unlike unsupervised non-learning hashing methods, BGE-based hashing starts from contextual dense embeddings, which can preserve semantic relationships beyond lexical overlap. We report three BGE model scales (small, base, and large), two similarity measures (cosine and Manhattan), and two hash quantization strategies: \textit{BGE-BIHash} and \textit{BGE-LSHash}. The results on CSFCUBE and RELISH are summarized in Tables~\ref{tab:csfcube_bge_results} and~\ref{tab:relish_bge_results}, respectively.

\begin{table*}[!htbp]
    \centering
    \small
    \renewcommand\arraystretch{1.2}
    \setlength{\tabcolsep}{1.8mm}
    \caption{Semantic-sensitive BGE-based hashing results on the CSFCUBE dataset. MAP and NDCG@20 are reported for background, method, result, and overall facets. Time denotes scoring time in seconds.}
    \label{tab:csfcube_bge_results}
    \begin{tabular}{llccccccccr}
        \toprule
        \multirow{2}{*}{\textbf{Method}} & \multirow{2}{*}{\textbf{Sim.}} &
        \multicolumn{4}{c}{\textbf{MAP}} & \multicolumn{4}{c}{\textbf{NDCG@20}} & \multirow{2}{*}{\textbf{Time}} \\
        \cmidrule(lr){3-6}\cmidrule(lr){7-10}
        & & Bg. & Meth. & Res. & All & Bg. & Meth. & Res. & All & \\
        \midrule
        BGE(small) & cosine & 0.4821 & 0.2292 & 0.2715 & 0.3245 & 0.7027 & 0.4220 & 0.5084 & 0.5412 & 40.1 \\
        BGE-BIHash(small) & cosine & 0.3792 & 0.1789 & 0.2266 & 0.2592 & 0.6131 & 0.3371 & 0.4380 & 0.4597 & 38.5 \\
        BGE-LSHash(small) & cosine & 0.3350 & 0.2023 & 0.2134 & 0.2486 & 0.5766 & 0.3852 & 0.4127 & 0.4558 & 101.2 \\
        BGE(small) & Manhattan & 0.4757 & 0.2261 & 0.2783 & 0.3237 & 0.6970 & 0.4182 & 0.5160 & 0.5407 & 41.2 \\
        BGE-BIHash(small) & Manhattan & 0.3891 & 0.2093 & 0.2320 & 0.2746 & 0.5954 & 0.2997 & 0.4189 & 0.4349 & 35.3 \\
        BGE-LSHash(small) & Manhattan & 0.3284 & 0.2005 & 0.2267 & 0.2503 & 0.5818 & 0.3899 & 0.4097 & 0.4580 & 130.6 \\
        BGE(base) & cosine & 0.5195 & 0.2260 & 0.2790 & 0.3379 & 0.7209 & 0.4248 & 0.5279 & 0.5546 & 42.8 \\
        BGE-BIHash(base) & cosine & 0.3390 & 0.1960 & 0.2248 & 0.2516 & 0.5814 & 0.3476 & 0.4415 & 0.4543 & 44.7 \\
        BGE-LSHash(base) & cosine & 0.4019 & 0.1805 & 0.2248 & 0.2664 & 0.6071 & 0.3510 & 0.4299 & 0.4597 & 97.2 \\
        BGE(base) & Manhattan & 0.5233 & 0.2231 & 0.2797 & 0.3384 & 0.7299 & 0.4220 & 0.5353 & 0.5591 & 45.8 \\
        BGE-BIHash(base) & Manhattan & 0.3232 & 0.1732 & 0.2141 & 0.2351 & 0.5472 & 0.3554 & 0.4031 & 0.4330 & 37.2 \\
        BGE-LSHash(base) & Manhattan & 0.4072 & 0.1787 & 0.2262 & 0.2680 & 0.6045 & 0.3489 & 0.4244 & 0.4563 & 74.3 \\
        BGE(large) & cosine & 0.4849 & 0.2344 & 0.3088 & 0.3399 & 0.7208 & 0.4467 & 0.5649 & 0.5746 & 68.1 \\
        BGE-BIHash(large) & cosine & 0.3196 & 0.1420 & 0.2127 & 0.2229 & 0.5527 & 0.3078 & 0.4332 & 0.4288 & 82.6 \\
        BGE-LSHash(large) & cosine & 0.3634 & 0.1615 & 0.2747 & 0.2646 & 0.6077 & 0.3516 & 0.4570 & 0.4694 & 245.3 \\
        BGE(large) & Manhattan & 0.4916 & 0.2337 & 0.3126 & 0.3431 & 0.7227 & 0.4495 & 0.5625 & 0.5754 & 69.9 \\
        BGE-BIHash(large) & Manhattan & 0.3221 & 0.1284 & 0.2082 & 0.2175 & 0.5524 & 0.2486 & 0.4198 & 0.4040 & 68.3 \\
        BGE-LSHash(large) & Manhattan & 0.3671 & 0.1680 & 0.2821 & 0.2705 & 0.6201 & 0.3587 & 0.4546 & 0.4750 & 169.1 \\
        \bottomrule
    \end{tabular}
\end{table*}

\begin{table*}[!htbp]
    \centering
    \small
    \renewcommand\arraystretch{1.2}
    \setlength{\tabcolsep}{2.2mm}
    \caption{Semantic-sensitive BGE-based hashing results on the RELISH dataset. MAP and NDCG@20 are reported on the test and development splits. Time denotes scoring time in seconds.}
    \label{tab:relish_bge_results}
    \begin{tabular}{llccccr}
        \toprule
        \multirow{2}{*}{\textbf{Method}} & \multirow{2}{*}{\textbf{Sim.}} &
        \multicolumn{2}{c}{\textbf{MAP}} & \multicolumn{2}{c}{\textbf{NDCG@20}} & \multirow{2}{*}{\textbf{Time}} \\
        \cmidrule(lr){3-4}\cmidrule(lr){5-6}
        & & Test & Dev & Test & Dev & \\
        \midrule
        BGE(small) & cosine & 0.6459 & 0.6533 & 0.8092 & 0.8150 & 1559.3 \\
        BGE-BIHash(small) & cosine & 0.5506 & 0.5562 & 0.7247 & 0.7326 & 1926.9 \\
        BGE-LSHash(small) & cosine & 0.4912 & 0.4983 & 0.6614 & 0.6695 & 2828.5 \\
        BGE(small) & Manhattan & 0.6459 & 0.6527 & 0.8090 & 0.8150 & 1554.0 \\
        BGE-BIHash(small) & Manhattan & 0.5350 & 0.5414 & 0.7092 & 0.7188 & 1847.6 \\
        BGE-LSHash(small) & Manhattan & 0.5389 & 0.5447 & 0.7121 & 0.7182 & 2682.7 \\
        BGE(base) & cosine & 0.6532 & 0.6608 & 0.8172 & 0.8227 & 6045.7 \\
        BGE-BIHash(base) & cosine & 0.5363 & 0.5466 & 0.7120 & 0.7222 & 1952.6 \\
        BGE-LSHash(base) & cosine & 0.4878 & 0.4925 & 0.6577 & 0.6697 & 2857.5 \\
        BGE(base) & Manhattan & 0.6540 & 0.6592 & 0.8169 & 0.8211 & 1800.1 \\
        BGE-BIHash(base) & Manhattan & 0.5216 & 0.5312 & 0.6969 & 0.7087 & 1884.5 \\
        BGE-LSHash(base) & Manhattan & 0.5323 & 0.5380 & 0.7051 & 0.7175 & 2727.8 \\
        BGE(large) & cosine & 0.6627 & 0.6652 & 0.8239 & 0.8268 & 3274.0 \\
        BGE-BIHash(large) & cosine & 0.5496 & 0.5476 & 0.7231 & 0.7273 & 3403.4 \\
        BGE-LSHash(large) & cosine & 0.4889 & 0.4971 & 0.6623 & 0.6763 & 4707.5 \\
        BGE(large) & Manhattan & 0.6626 & 0.6650 & 0.8242 & 0.8271 & 3198.5 \\
        BGE-BIHash(large) & Manhattan & 0.5288 & 0.5285 & 0.7032 & 0.7080 & 3300.8 \\
        BGE-LSHash(large) & Manhattan & 0.5336 & 0.5416 & 0.7114 & 0.7231 & 4692.8 \\
        \bottomrule
    \end{tabular}
\end{table*}

The CSFCUBE results show that dense BGE embeddings provide the strongest semantic-sensitive performance in this setting. The best overall CSFCUBE result is achieved by BGE(large) with Manhattan similarity (0.3431 MAP, 0.5754 NDCG@20). The background facet is generally easier than the method facet, which is consistent with the intuition that topical relatedness is easier to preserve than fine-grained methodological similarity. Quantized variants (BGE-BIHash and BGE-LSHash) reduce storage or comparison complexity, but they also reduce retrieval quality relative to dense BGE. On RELISH, test and development results are close across most BGE configurations, indicating stable split-level behavior. Dense BGE again performs best, with BGE(large)+Manhattan reaching 0.6626 MAP and 0.8242 NDCG@20 on the test split. Compact BGE-BIHash and BGE-LSHash remain below the dense baseline, reinforcing the compression-versus-quality trade-off. Rows with atypical values are treated as data-consistency checks, and the interpretation emphasizes trends across metrics and datasets.

\subsection{Semantic Robustness under Text Compression (RQ3)}

\begin{figure*}[htbp]
    \centering
    \captionsetup{font=footnotesize}
    \captionsetup[subfigure]{font=scriptsize, skip=0.2pt}
    
    \begin{subfigure}{0.4\linewidth}
        \centering
        \includegraphics[width=0.95\linewidth]{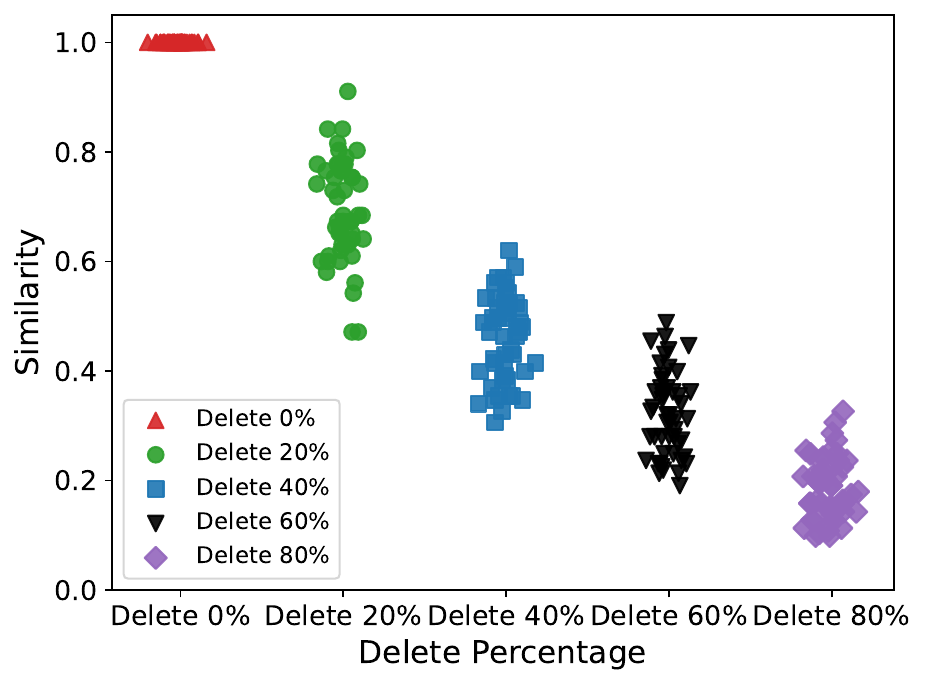}
        \caption{MinHash}
        \label{subfig:minhash}
    \end{subfigure}
    \hspace{0.2mm}
    \begin{subfigure}{0.4\linewidth}
        \centering
        \includegraphics[width=0.95\linewidth]{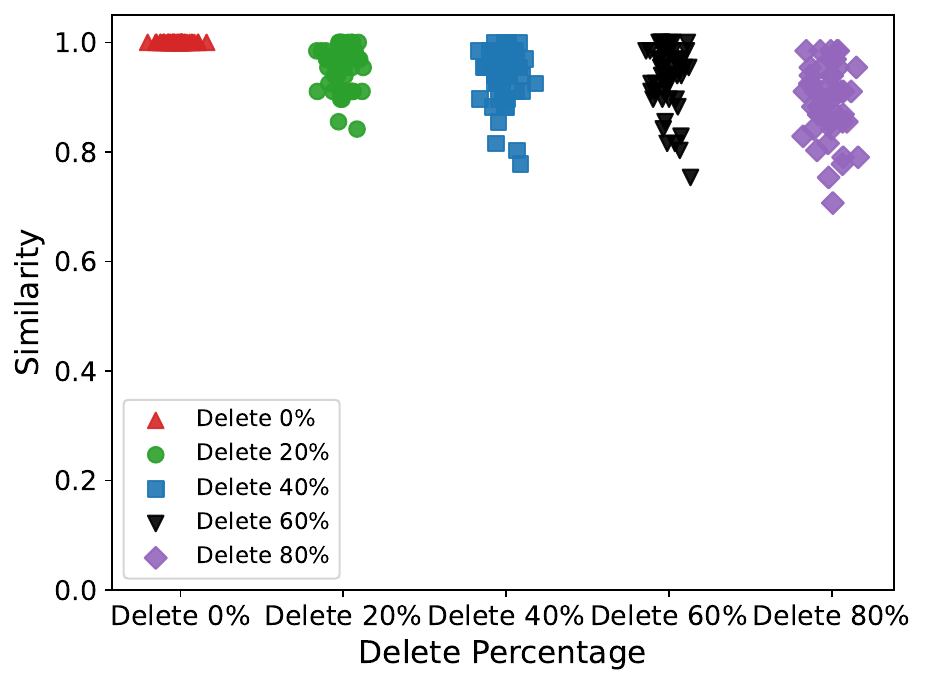}
        \caption{SimHash}
        \label{subfig:simhash}
    \end{subfigure}

    \vspace{0mm}

    \begin{subfigure}{0.4\linewidth}
        \centering
        \includegraphics[width=0.95\linewidth]{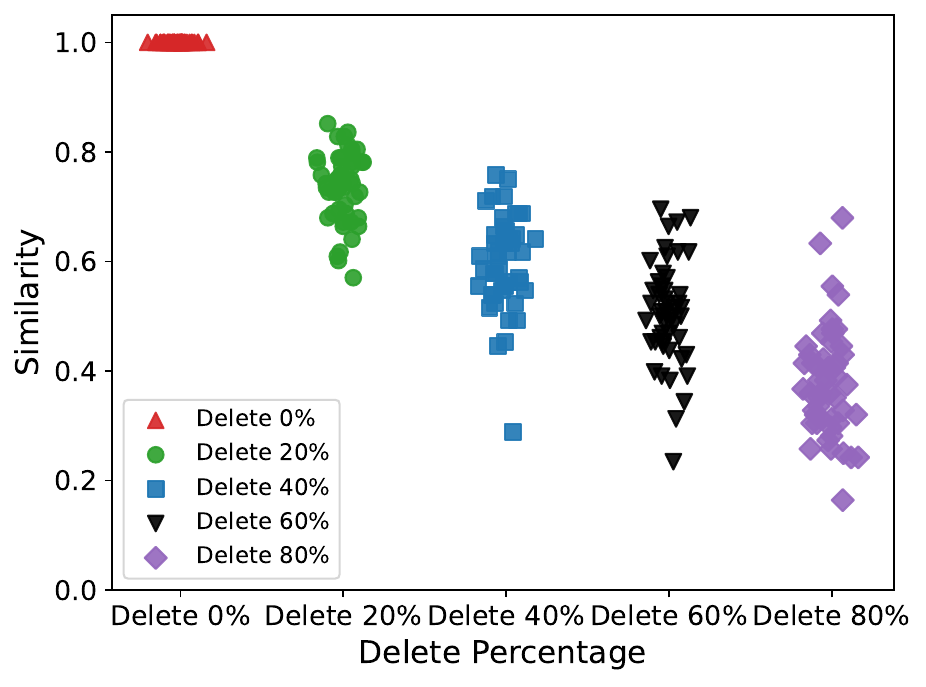}
        \caption{FlyHash}
        \label{subfig:flyhash}
    \end{subfigure}
    \hspace{0.2mm}
    \begin{subfigure}{0.4\linewidth}
        \centering
        \includegraphics[width=0.95\linewidth]{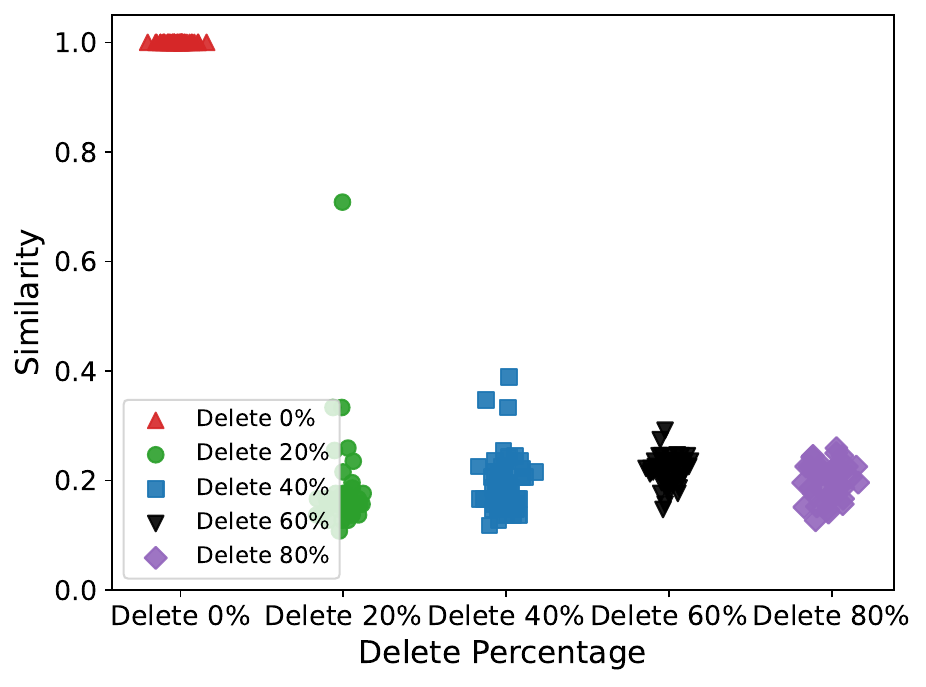}
        \caption{FuzzyHash}
        \label{subfig:fuzzyhash}
    \end{subfigure}

    \vspace{0mm}

    \begin{subfigure}{0.4\linewidth}
        \centering
        \includegraphics[width=0.95\linewidth]{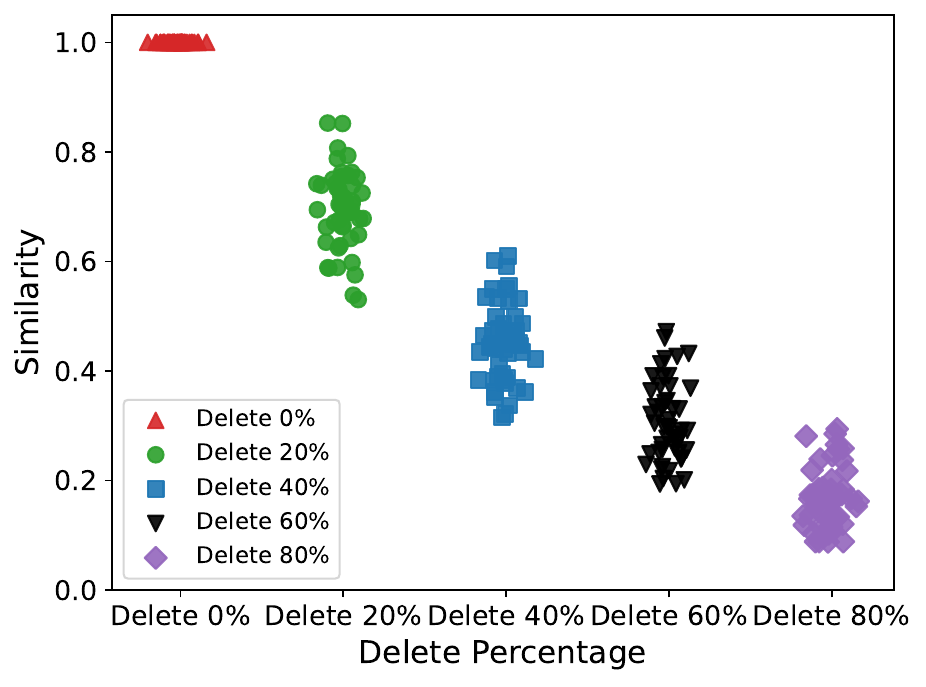}
        \caption{Winnowing}
        \label{subfig:winnowing}
    \end{subfigure}
    \hspace{0.2mm}
    \begin{subfigure}{0.4\linewidth}
        \centering
        \includegraphics[width=0.95\linewidth]{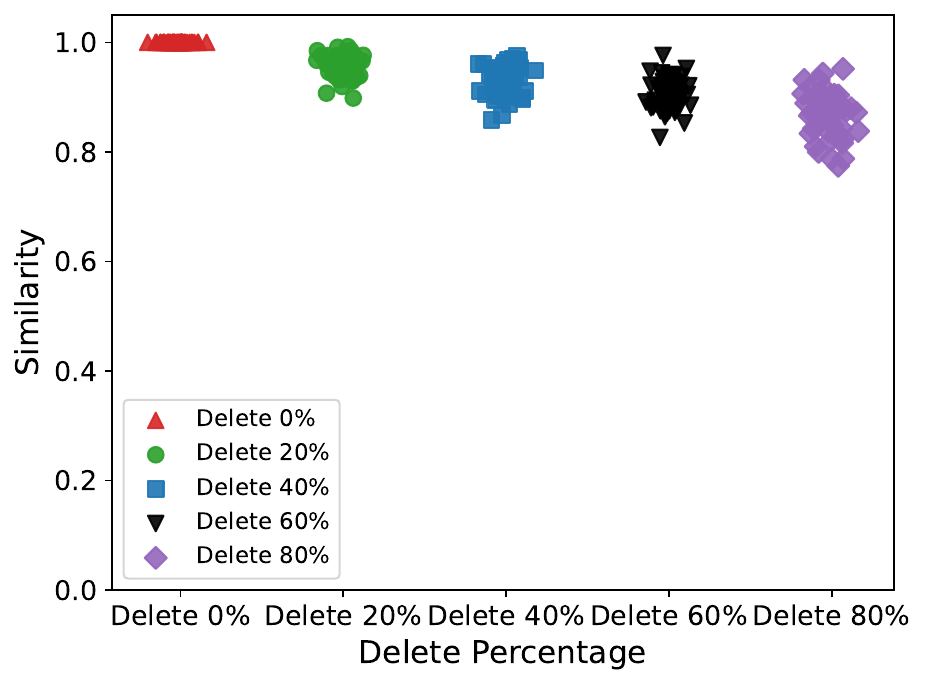}
        \caption{BGE-BIHash}
        \label{subfig:bgebinhash}
    \end{subfigure}

    \vspace{0mm}

    \begin{subfigure}{0.4\linewidth}
        \centering
        \includegraphics[width=0.95\linewidth]{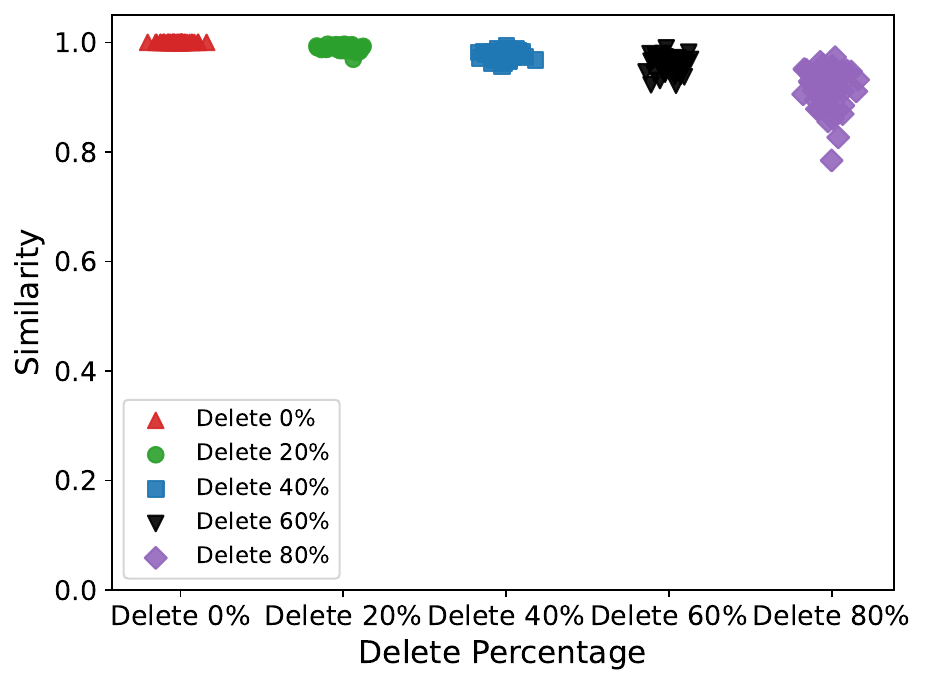}
        \caption{BGE-base}
        \label{subfig:bgehash}
    \end{subfigure}
    \hspace{0.2mm}
    \begin{subfigure}{0.4\linewidth}
        \centering
        \includegraphics[width=0.95\linewidth]{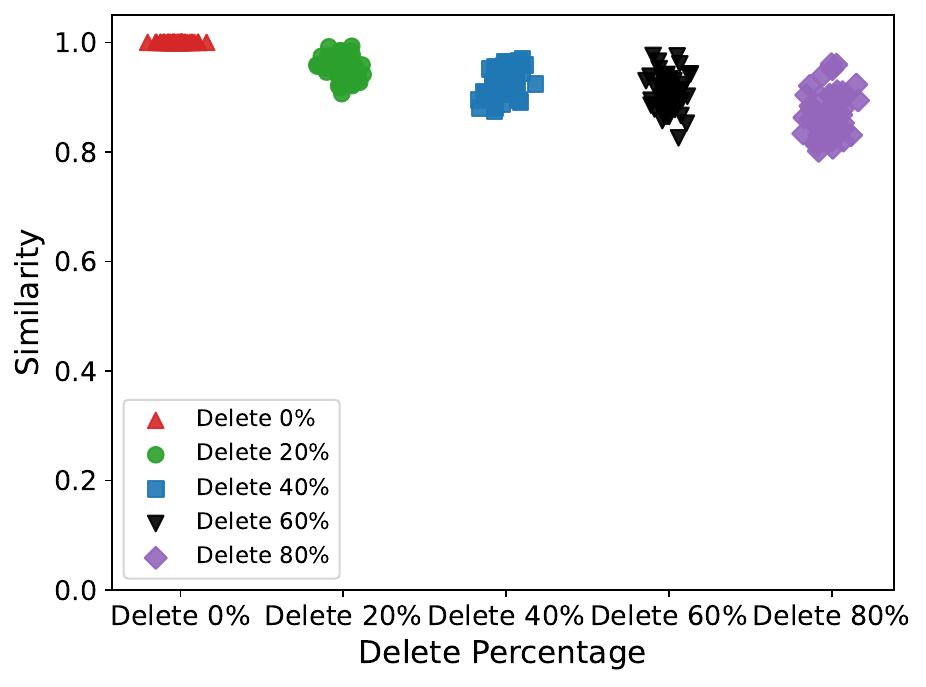}
        \caption{BGE-LSHash}
        \label{subfig:bgelshhash}
    \end{subfigure}

    \caption{
        \textbf{Comparison of Semantic-Agnostic and Semantic-Sensitive Hashing Robustness on the CSFCUBE Dataset.}
        The subfigures compare similarity changes under progressive text deletion for semantic-agnostic methods and BGE-based semantic-sensitive methods. Mapping: (a) MinHash, (b) SimHash, (c) FlyHash, (d) FuzzyHash, (e) Winnowing, (f) BGE-BIHash, (g) BGE-base, (h) BGE-LSHash.
    }
    \label{fig:hashing_delete_comparison}
\end{figure*}

To investigate how different hashing paradigms preserve semantic information under textual compression, we selected 50 documents from the \textbf{CSFCUBE} dataset and progressively shortened them using an LLM-based summarization process. Figure~\ref{fig:hashing_delete_comparison} presents the variation of similarity values across five deletion ratios (0\%, 20\%, 40\%, 60\%, and 80\%) for both \textit{semantic-sensitive} and \textit{semantic-agnostic} hashing methods. This experiment is not intended to simulate all possible edits; instead, it provides a controlled probe of how each representation responds when lexical evidence is removed while the main topic is partially preserved.

The results reveal a clear contrast between surface-oriented fingerprints and BGE-based representations under text compression. The semantic-agnostic methods generally show decreasing similarity as the deletion ratio increases, while the BGE-based methods retain higher similarity scores under moderate compression. This suggests that semantic-sensitive representations are better suited to matching summarized or paraphrased scientific abstracts, whereas semantic-agnostic methods are more appropriate when lexical overlap is expected to remain high.

Overall, these findings indicate a practical distinction between semantic-agnostic and semantic-sensitive hashing paradigms: surface-based fingerprints are efficient for near-duplicate detection, whereas BGE-based methods are more robust when document content is compressed or rewritten.

\subsection{Effect of the Number of Hash Tables}
To evaluate the influence of the number of hash tables on the BGE-LSHash framework, we varied $k$ from 4 to 16 while keeping the hash length $L \approx 128$ fixed. As shown in Table~\ref{tab:lsh_bge_tables}, both MAP and NDCG remain nearly constant (within $\pm$1\%) across different $k$ values, suggesting that retrieval performance is largely insensitive to the number of hash tables in this setting. The underlying BGE embeddings may already form a well-structured semantic space, where similar documents are mapped closely enough that additional hash tables contribute little to recall improvement. The end-to-end scoring time is also generally stable across most settings, with $k=16$ showing a modest increase. This pattern indicates that the measured runtime is dominated by BGE-based fingerprint generation rather than by the LSH table lookup alone; the latter is expected to scale approximately linearly with $k$, but its contribution is relatively small in the full pipeline. Therefore, smaller $k$ values (e.g., $k=4$ or $k=8$) offer an efficient balance between retrieval accuracy and computational cost for the evaluated RELISH configuration.

\begin{table*}[htbp]
\caption{Impact of the number of hash tables ($k$) on BGE-LSHash performance on RELISH. MAP corresponds to average precision on the test split. Scoring time is the end-to-end scoring wall time reported in the logs; FP time measures query and candidate fingerprint generation, while Similarity time measures pairwise similarity.}
\renewcommand\arraystretch{1.4}
\centering
\resizebox{0.88\textwidth}{!}{%
\begin{tabular}{lcccccc}
\toprule
\textbf{Model} & \makecell{\textbf{Hash tables ($k$)}} & \textbf{MAP} & \textbf{NDCG@20} & \makecell{\textbf{Scoring time (s)}} & \makecell{\textbf{FP time (s)}} & \makecell{\textbf{Similarity time (s)}} \\
\midrule
BGE-LSHash & 4  & 0.5401 & 0.7143 & 51,356.47 & 51,238.39 & 116.11 \\
BGE-LSHash & 6  & 0.5310 & 0.7054 & 51,575.09 & 51,455.62 & 117.41 \\
BGE-LSHash & 8  & 0.5380 & 0.7144 & 49,252.63 & 49,141.67 & 108.92 \\
BGE-LSHash & 12 & 0.5322 & 0.7055 & 51,545.08 & 51,423.72 & 119.33 \\
BGE-LSHash & 16 & 0.5330 & 0.7111 & 54,681.65 & 54,546.51 & 132.96 \\
\bottomrule
\end{tabular}
}
\label{tab:lsh_bge_tables}
\end{table*}

\subsection{Threats to Validity}
Several factors may affect the generality of the results.
\begin{itemize}[leftmargin=2em]
    \item \textbf{Dataset scope}: CSFCube is facet-rich but small, while RELISH is larger but lacks facet labels. Conclusions may shift on larger multi-domain corpora or on domains with different writing conventions.
    \item \textbf{Metric--representation coupling}: some scorers become rank-equivalent for specific representations, which can mask apparent diversity of metric choices. We report representative rows where appropriate, but this behavior should be rechecked when implementations or output formats change.
    \item \textbf{Runtime portability}: reported times are useful for relative comparison within this implementation, but absolute values may change across hardware and software stacks. The timing results should therefore be read together with the qualitative cost patterns rather than as universal throughput claims.
\end{itemize}

\subsection{Benchmark Positioning and Novelty}
H3D contributes an evaluation protocol rather than a new neural architecture. The benchmark combines three elements: (i) a unified task definition with fixed query--candidate inputs and scorer-aligned hash comparison, (ii) joint evaluation of semantic-agnostic and semantic-sensitive hashing under the same ranking pipeline, and (iii) robustness analysis under controlled text compression.

These elements keep the comparison consistent across hashing families and make it easier to separate three questions that are often conflated: how a fingerprint is generated, how two fingerprints are compared, and how much semantic information survives compression. The reported results are conditional on the stated datasets, scorers, and implementations.

\section{Conclusion and Future Work}

H3D provides an unsupervised text hashing benchmark for fine-grained document deduplication. By evaluating unsupervised non-learning and semantic-sensitive hashing methods under a unified protocol on CSFCube and RELISH, we characterized how representation type and scorer choice jointly affect retrieval quality, robustness, and efficiency. Results suggest that rule-based lexical and structural hashes remain practical for near-duplicate retrieval, whereas BGE-based semantic hashing is more robust under meaning-preserving compression and rewriting, at higher computational cost. The benchmark also shows that similarity functions cannot be treated as interchangeable: for some representations they are rank-equivalent, while for others they substantially change the ordering of candidate documents.

Future work will expand H3D with broader domains, larger corpora, and stronger perturbation families, such as paraphrase, translation, and citation edits. We also plan to study learned or calibration-aware quantizers on top of dense encoders, with explicit neighborhood-preservation objectives and uncertainty-aware evaluation.

\appendix
\section{Additional Reproducibility and Diagnostic Notes}

\begin{figure*}[!htbp]
    \centering
    \includegraphics[width=0.98\textwidth]{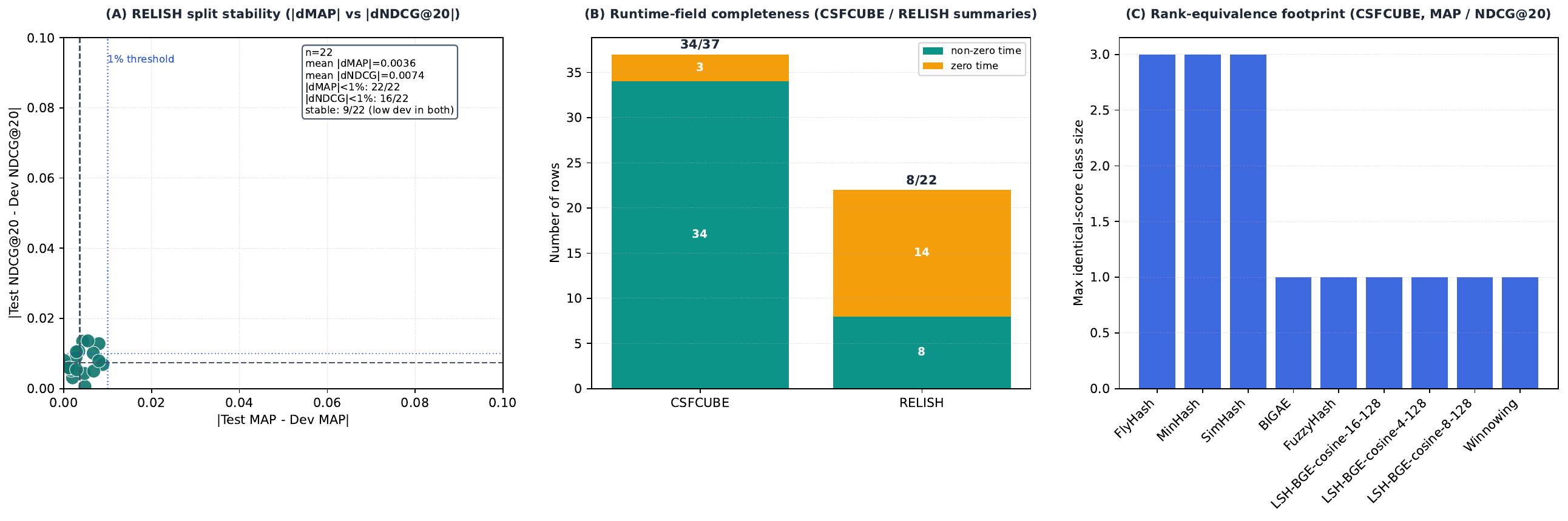}
    \caption{Appendix diagnostics from released summary artifacts. (A) RELISH split-gap stability across configurations using absolute Test--Dev differences in MAP and NDCG@20. (B) Runtime-field completeness check in the released CSFCUBE/RELISH summary CSV files. (C) Maximum identical-score class size per hash family on CSFCUBE using MAP/NDCG@20 pairs.}
    \label{fig:appendix_diagnostics_overview}
\end{figure*}

\subsection{Split Stability Statistics on RELISH}
Figure~\ref{fig:appendix_diagnostics_overview}(A) visualizes absolute test--dev gaps for all 22 configurations in \textit{relish\_summary.csv}. The average absolute gap is $0.0036$ for MAP and $0.0074$ for NDCG@20, while the maximum observed gaps are $0.0089$ (MAP) and $0.0136$ (NDCG@20). These small gaps indicate stable split-level ranking behavior in the current benchmark setting. A grouped view shows slightly larger MAP variation for LSH-BGE settings (mean $|\Delta\text{MAP}|=0.0060$, $n=5$) than for other methods (mean $|\Delta\text{MAP}|=0.0030$, $n=17$), but both remain small in absolute scale.

\subsection{Timing-Field Integrity Check in Released Summaries}
Figure~\ref{fig:appendix_diagnostics_overview}(B) summarizes a timing-field consistency check on the released CSV summaries to clarify runtime interpretation boundaries.
\begin{itemize}[leftmargin=2em]
    \item In \textit{csfcube\_summary.csv}, 3 out of 37 rows have zero runtime values.
    \item In \textit{relish\_summary.csv}, 14 out of 22 rows have zero runtime values.
\end{itemize}
Inspection of the summarization scripts indicates a likely parsing mismatch: the RELISH summarizer accumulates time from lines matching \texttt{Finished all scoring in ...}, while many logs report \texttt{Finished scoring in ...}. Runtime tables are therefore comparative within the released summaries, with potential missing entries where log formats differ.

\subsection{Data-Consistency Notes for Aggregated Artifacts}
Figure~\ref{fig:appendix_diagnostics_overview}(C) provides a compact view of rank-equivalence footprint by reporting, for each hash family, the largest identical-score class size observed in CSFCUBE summaries.
Each reported row is linked to one canonical source path in the released artifacts, consisting of the configuration file, run log, and aggregated file.

\begin{figure*}[!htbp]
    \centering
    \includegraphics[width=0.78\textwidth]{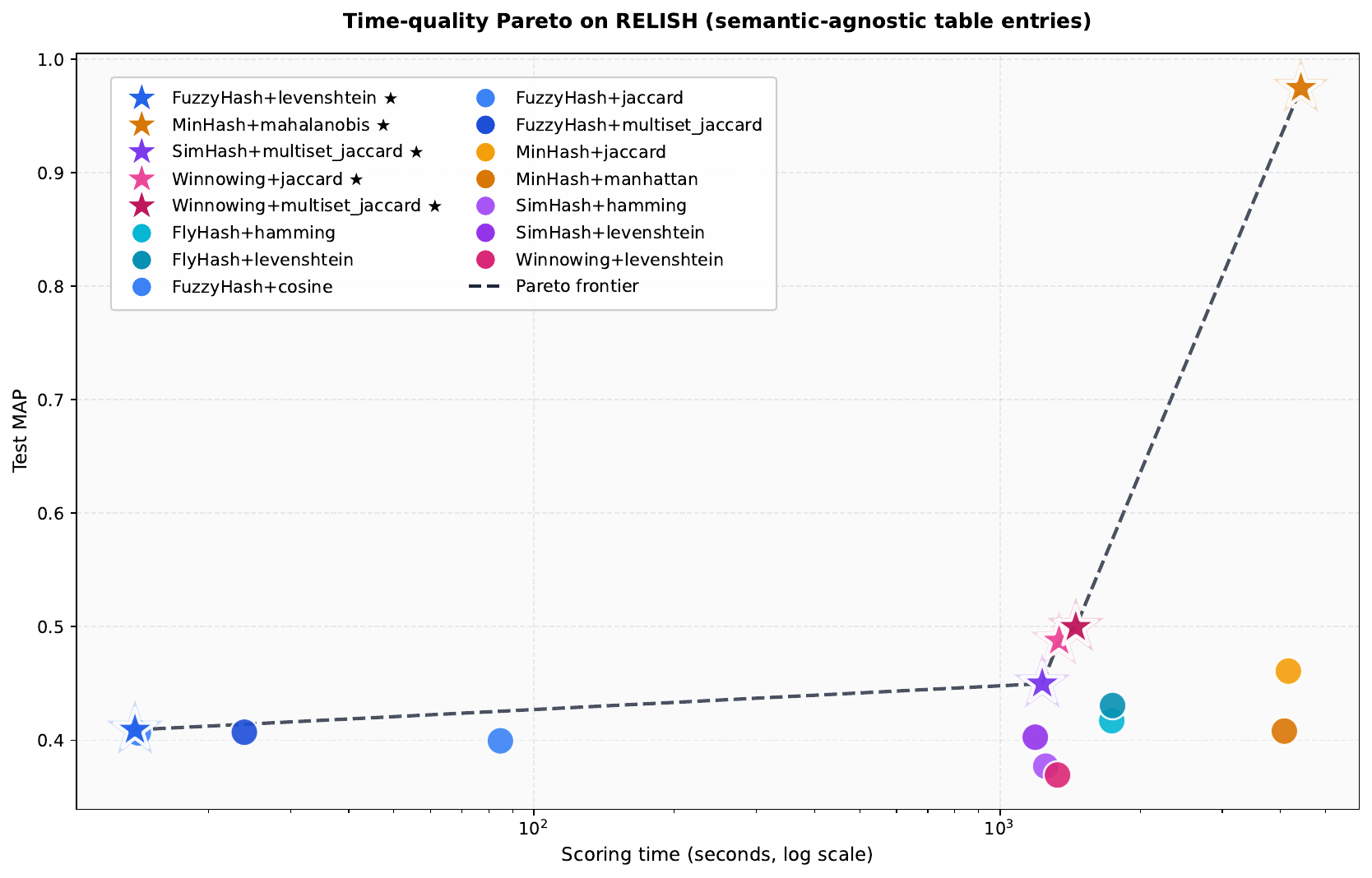}
    \caption{Time--quality Pareto scatter on RELISH semantic-agnostic configurations from Table~\ref{tab:relish_results} (x-axis: scoring time in log scale; y-axis: Test MAP). Starred points and dashed line indicate Pareto-optimal trade-off candidates.}
    \label{fig:appendix_time_quality_pareto}
\end{figure*}

\subsection{Time--Quality Pareto View}
Figure~\ref{fig:appendix_time_quality_pareto} complements Table~\ref{tab:relish_results} by plotting RELISH semantic-agnostic configurations in a time--quality plane. The starred configurations indicate Pareto-optimal candidates under the reported scoring time and Test MAP, making the accuracy--efficiency trade-off easier to inspect than in the table alone.

\subsection{Additional Metric Coverage from Aggregated Evaluations}
\begin{figure*}[!htbp]
    \centering
    \includegraphics[width=0.90\textwidth]{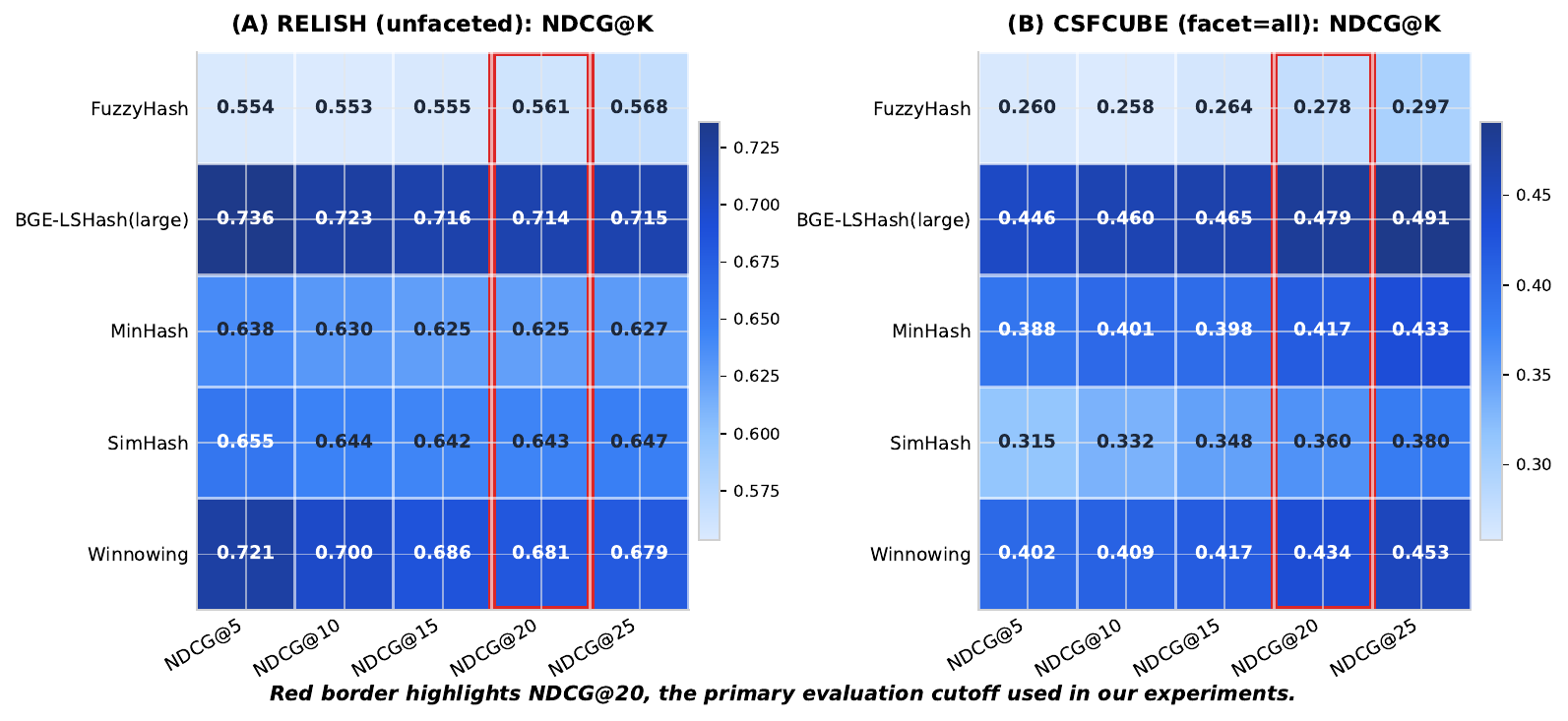}
    \caption{NDCG@K performance across different K values on RELISH (unfaceted) and CSFCUBE (facet=all) datasets. All results were obtained by averaging over three independent runs, each conducted on a 10\% random sample of the development (dev) set.}
    \label{fig:appendix_extended_metrics_profile_heatmap_v4.pdf}
\end{figure*}

\begin{figure*}[!htbp]
    \centering
    \includegraphics[width=0.90\textwidth]{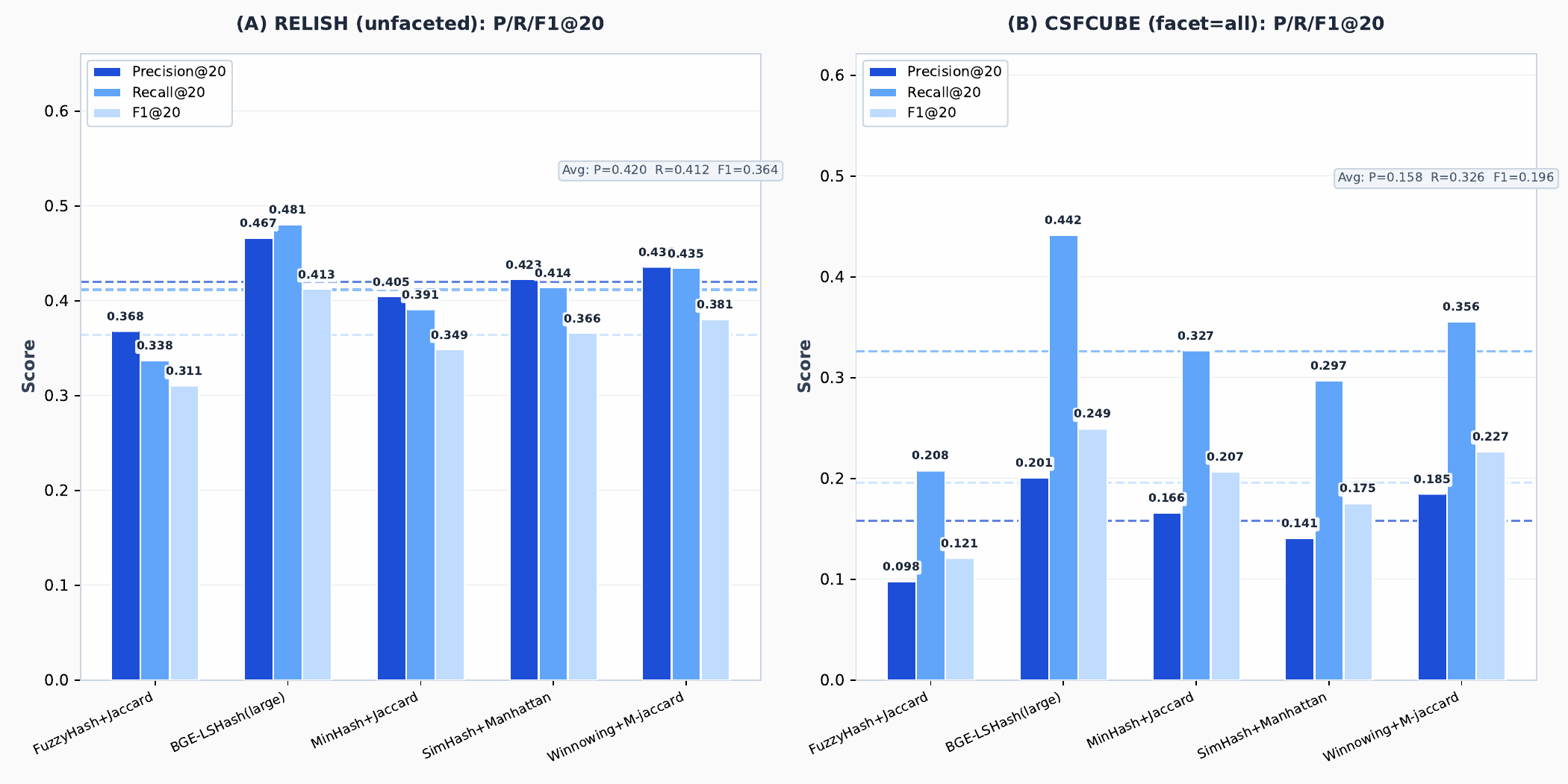}
    \caption{Extended metrics from aggregated evaluations for representative methods (Winnowing+M-jaccard, SimHash+Manhattan, MinHash+Jaccard, FuzzyHash+Jaccard, and LSH-BGE(8,128)). (A) RELISH test-unfaceted NDCG@K; (B) CSFCUBE test-facet=all NDCG@K; (C) RELISH precision/recall/F1@20; (D) CSFCUBE precision/recall/F1@20.}
    \label{fig:appendix_extended_metrics_profile_heatmap_v4_bar}
\end{figure*}

Current main tables focus on MAP and NDCG@20 for comparability, but the released aggregated files also contain richer ranking metrics (NDCG@5/10/15/20/25, precision@K, recall@K, and F1@K). Figure~\ref{fig:appendix_extended_metrics_profile_heatmap_v4.pdf} visualizes representative NDCG@K profiles from both semantic-agnostic and BGE-based settings on two datasets, while Figure~\ref{fig:appendix_extended_metrics_profile_heatmap_v4_bar} provides a companion panel view that also includes precision@20, recall@20, and F1@20.

In Figure~\ref{fig:appendix_extended_metrics_profile_heatmap_v4.pdf}, the NDCG@K trends show whether conclusions based on NDCG@20 are stable across smaller and larger cutoffs. Figure~\ref{fig:appendix_extended_metrics_profile_heatmap_v4_bar} then summarizes the same representative methods in a more compact bar-style view: panels (A) and (B) compare NDCG@K on RELISH (test, \textit{unfaceted}) and CSFCUBE (test, \textit{facet=all}), respectively, while panels (C) and (D) compare precision@20, recall@20, and F1@20 on the same dataset settings with explicit value labels.

\begin{table}[htbp]
    \centering
    \small
    \renewcommand\arraystretch{1.15}
    \setlength{\tabcolsep}{2.0mm}
    \caption{Documents used in the case study. Doc B is a near-duplicate paraphrase of Doc A, while Doc C is unrelated.}
    \resizebox{\linewidth}{!}{
    \begin{tabular}{lp{0.78\linewidth}}
        \toprule
        \textbf{Document} & \textbf{Text} \\
        \midrule
        Doc A (Original) & Machine learning is an important branch of artificial intelligence. It enables computers to automatically learn and improve through data-driven paradigms. \\
        Doc B (Near duplicate) & Machine learning belongs to a key subfield of artificial intelligence. It leverages data-driven methods to allow computers to self-learn and refine performance. \\
        Doc C (Unrelated) & Quantum computing represents an innovative computing paradigm built on quantum mechanics, utilizing quantum superposition and entanglement to achieve parallel computation. \\
        \bottomrule
    \end{tabular}}
    \label{tab:case_study_docs}
\end{table}

\section{Case study}

This section provides concrete, small-scale illustrations of how representative hashing pipelines transform document text into fingerprints and then compute similarity scores. Rather than replacing the benchmark tables, these examples make the intermediate representations explicit and help explain why different similarity functions behave differently across hash families.

We construct a controlled three-document case to distinguish near-duplicate matching from unrelated-document rejection. Table~\ref{tab:case_study_docs} defines the example inputs: Doc A is the original text, Doc B is a near-duplicate with light paraphrasing, and Doc C is topically unrelated. Based on this case, we set up five experiments by selecting strong representative combinations from each non-learning hash family: \textit{FlyHash+Jaccard}, \textit{MinHash+Jaccard}, \textit{SimHash+Manhattan}, \textit{Winnowing Hash+Jaccard}, and \textit{FuzzyHash+Levenshtein}. The goal is to test whether each pipeline assigns a high similarity score to the near-duplicate pair A--B while assigning a low score to the unrelated pair A--C.

Table~\ref{Indicators-Summary} complements the five examples above by summarizing discrimination strength and computational cost under a controlled one-positive/one-negative setting. FlyHash is the fastest configuration and achieves complete separation of the irrelevant sample, but its sparse activation pattern remains sensitive to small lexical perturbations. MinHash yields the strongest set-overlap fidelity and a high near-duplicate score, at the cost of heavier repeated hash computation. SimHash is compact and efficient, with clear bit-level separation between relevant and irrelevant pairs, although the Manhattan-score mapping compresses the numerical similarity range.

Winnowing also drives the irrelevant-pair similarity to zero and is therefore highly pure for local overlap detection, but it is the slowest configuration because of its window-scanning procedure. FuzzyHash produces the highest positive-pair similarity, indicating strong tolerance to local insertions, deletions, and substitutions; however, its negative-pair similarity remains relatively high, which weakens class separation. These observations are consistent with the case-study fingerprints themselves: methods that fit overlap or edit behavior more closely tend to incur higher computational cost, whereas faster signatures generally sacrifice either stability or score resolution.

All five pipelines reach MAP = 1.0 in this simplified setting, so each can separate the single positive pair from the single negative pair. The more informative comparison is therefore relative: FlyHash and SimHash are the most efficient; MinHash and FuzzyHash are strongest on the positive pair; Winnowing and FlyHash are strongest on negative-pair purity. The case study thus reinforces the benchmark-level conclusion that computation, overlap fidelity, and clean negative separation cannot be optimized simultaneously by a single hashing architecture.

\begin{expbox}{Experiment 1: FlyHash + Jaccard Similarity}
\stepitem{1}Construct a frequency vector over all 3-shingle features extracted from the preprocessed document, with dimension $D=216$.\\[4pt]
\stepitem{2}Load a pre-generated fixed random binary projection matrix $\mathbf{W} \in \{0,1\}^{216 \times 128}$, producing a 128-dimensional intermediate vector.\\[4pt]
\stepitem{3}Perform matrix multiplication between the original feature vector and the projection matrix to obtain a 128-dimensional floating-point mapped vector.\\[4pt]
\stepitem{4}Apply WTA (Winner-Take-All) selection: set the top 8 indices with the largest values to 1, and all remaining positions to 0.\\[4pt]
\stepitem{5}Generate the 128-bit sparse binary FlyHash signature.\\[4pt]
\stepitem{6}Extract the index sets where the hash value is 1 for both documents, then compute Jaccard similarity as the ratio of intersection size to union size.\\[8pt]
\textbf{Hash Output} (128-bit binary arrays, only indices with value 1 are listed, others are 0):\\[4pt]
\par\centering
\begin{hashoutputbox}
\small\texttt{%
\textbf{Hash(A)} active indices: \{4,11,16,27,35,42,59,73\}\\
Binary sequence segments:\\
0 1 0 0 1 0 0 0 0 0 0 1 0 0 0 0 1 0 0 0 0 0 0 0 $\cdots\cdots$ 0 0 0 1 0 0 0 0\\
0 0 1 0 0 0 0 0 1 0 0 0 0 0 0 0 0 0 0 0 0 0 0 0 $\cdots\cdots$ 0 0 0 0 0 0 0 0\\
0 0 0 0 0 0 0 0 0 0 1 0 0 0 0 0 0 0 0 0 0 0 0 0 $\cdots\cdots$ 0 0 0 0 0 0 0 0\\
0 0 0 0 0 0 0 0 0 0 0 0 0 0 0 0 0 0 0 0 0 0 0 0 $\cdots\cdots$ 0 0 0 0 0 0 0 0\\
(8 ones, 120 zeros)\\[4pt]
\textbf{Hash(B)} active indices: \{4,11,16,27,35,42,59,68\}\\
Binary sequence segments:\\
0 1 0 0 1 0 0 0 0 0 0 1 0 0 0 0 1 0 0 0 0 0 0 0 $\cdots\cdots$ 0 0 0 1 0 0 0 0\\
0 0 1 0 0 0 0 0 1 0 0 0 0 0 0 0 0 0 0 0 0 0 0 0 $\cdots\cdots$ 0 1 0 0 0 0 0 0\\
0 0 0 0 0 0 0 0 0 0 0 0 0 0 0 0 0 0 0 0 0 0 0 0 $\cdots\cdots$ 0 0 0 0 0 0 0 0\\
0 0 0 0 0 0 0 0 0 0 0 0 0 0 0 0 0 0 0 0 0 0 0 0 $\cdots\cdots$ 0 0 0 0 0 0 0 0\\
\textbf{Hash(C)} active indices: \{2,9,22,47,62,81,95,117\}\\
Binary sequence segments:\\
0 0 1 0 0 0 0 0 0 1 0 0 0 0 0 0 0 0 0 0 0 0 1 0 $\cdots\cdots$ 0 0 0 0 0 0 0 0\\
0 0 0 0 0 0 0 0 0 0 0 0 0 0 0 0 0 0 0 0 0 0 0 0 $\cdots\cdots$ 0 0 0 0 0 0 0 0\\
0 0 0 0 0 0 0 0 0 0 0 0 1 0 0 0 0 0 0 0 0 1 0 0 $\cdots\cdots$ 0 0 0 0 0 0 0 0\\
0 0 0 0 0 0 0 0 0 0 0 0 0 0 0 0 0 0 0 0 0 0 0 1 $\cdots\cdots$ 0 0 0 0 0 0 0 1\\
\textbf{Matching Statistics:} A and B overlap on 7 active positions, while A and C overlap on 1 active position.
}
\end{hashoutputbox}
\par
\end{expbox}

\begin{expbox}{Experiment 2: MinHash + Jaccard Similarity}
\stepitem{1}Map all 3-shingle features of the document into a set $S$ of unsigned integers.\\[4pt]
\stepitem{2}Predefine 100 independent random permutation hash functions $h_1, h_2, \ldots, h_{100}$.\\[4pt]
\stepitem{3}For a single document, iterate over the feature set $S$, substitute each feature into every $h_i$, and record the minimum hash output for each group.\\[4pt]
\stepitem{4}Concatenate the 100 minimum values in the order of the hash functions to generate a 100-dimensional integer MinHash signature.\\[4pt]
\stepitem{5}Compare the signature arrays of two documents and count the total number of positions where the values are equal.\\[4pt]
\stepitem{6}Approximate Jaccard similarity = (number of matching equal positions) / (total number of hash functions 100).\\[8pt]
\textbf{Hash Output} (100-dimensional integer signatures, beginning and end fully shown, middle partially omitted):\\[4pt]
\par\centering
\begin{hashoutputbox}
\small\texttt{%
\textbf{MinHash(A):}\\\relax
[1245, 892, 3571, 629, 4813, 907, 2664, 711, 5208, 1036, 3142, 584, $\cdots\cdots$, 4962, 687, 2873, 858, 5086, 1051, 3895, 775]\\[4pt]
\textbf{MinHash(B):}\\\relax
[1245, 892, 3571, 629, 4813, 907, 2664, 711, 5208, 1036, 3142, 584, $\cdots\cdots$, 4962, 687, 2873, 858, 5086, 1051, 3895, 726]\\
\textbf{MinHash(C):}\\\relax
[7621, 4401, 1093, 6744, 8205, 3362, 917, 5540, 7126, 2893, 4055, 6618, $\cdots\cdots$, 1376, 5304, 8794]\\[4pt]
\textbf{Matching Statistics:} A and B have 92 matching positions; A and C have only 7 matching positions.
}
\end{hashoutputbox}
\par
\end{expbox}

\begin{expbox}{Experiment 3: SimHash + Manhattan Similarity}
\stepitem{1}Tokenize the text and assign a term-frequency weight to each token.\\[4pt]
\stepitem{2}Generate a conventional 64-bit binary hash for each individual token.\\[4pt]
\stepitem{3}Initialize a 64-dimensional floating-point vector $V$. For each token, update $V$ with weights: if the hash bit is 1, add the weight to the corresponding position in $V$; if 0, subtract the weight.\\[4pt]
\stepitem{4}Binarize $V$ to generate the SimHash signature: if $V[i] > 0$, set the $i$-th bit of the signature to 1; otherwise, set it to 0.\\[4pt]
\stepitem{5}Convert the binary SimHash into a 64-dimensional floating-point vector.\\[4pt]
\stepitem{6}Compute Manhattan distance $d = \sum |vA_i - vB_i|$; normalized similarity = $1/(1+d)$.\\[8pt]
\textbf{Hash Output} (64-bit complete binary, middle partially omitted, sufficient bits retained at both ends):\\[4pt]
\par\centering
\begin{hashoutputbox}
\small\texttt{%
\textbf{SimHash(A):} \codetext{01101011 10010100 00110110 $\cdots\cdots$ 01011001 1010}\\[4pt]
\textbf{SimHash(B):} \codetext{01101011 10010100 00110110 $\cdots\cdots$ 01011001 1000}\\[4pt]
\textbf{SimHash(C):} \codetext{10010100 01101011 11001001 $\cdots\cdots$ 10100110 0101}\\[4pt]
\textbf{Matching Statistics:} Compared with A, B differs in 2 bits, while C differs in 31 bits.
}
\end{hashoutputbox}
\par
\end{expbox}

\begin{expbox}{Experiment 4: Winnowing Hash + Jaccard Similarity}
\stepitem{1}Generate 3-shingles from the text and map each shingle to an integer hash value.\\[4pt]
\stepitem{2}Set the window size $w = 5$ and slide the window across the hash sequence.\\[4pt]
\stepitem{3}Within each window, select the minimum value as the window fingerprint; deduplicate to obtain the document's Winnowing fingerprint set.\\[4pt]
\stepitem{4}Compute the standard Jaccard similarity based on the fingerprint sets of two documents.\\[8pt]
\textbf{Hash Output} (window fingerprint sets, beginning and end fully shown, middle lightly omitted):\\[4pt]
\par\centering
\begin{hashoutputbox}
\small\texttt{%
\textbf{Winnow(A)} = \{123, 456, 789, 1024, 1357, $\cdots\cdots$, 1492, 1680\} (7 fingerprints in total)\\[4pt]
\textbf{Winnow(B)} = \{123, 456, 789, 1024, 1361, $\cdots\cdots$, 1501, 1702\}\\[4pt]
\textbf{Winnow(C)} = \{987, 654, 321, 2048, 2714, $\cdots\cdots$, 2963, 3001\}\\[4pt]
\textbf{Matching Statistics:} A and B share 5 fingerprints, while A and C share none.
}
\end{hashoutputbox}
\par
\end{expbox}

\begin{expbox}{Experiment 5: FuzzyHash + Levenshtein Similarity}
\stepitem{1}Partition the text into fixed-length data blocks and generate a rolling weak hash for each block.\\[4pt]
\stepitem{2}Concatenate the block hashes to produce a variable-length Base64 fuzzy fingerprint (FuzzyHash).\\[4pt]
\stepitem{3}Compute the Levenshtein edit distance (number of insertions, deletions, and substitutions) between two FuzzyHash strings.\\[4pt]
\stepitem{4}Normalized similarity = $1 - (\text{edit distance} / \max(\text{len(hashA)}, \text{len(hashB)}))$.\\[8pt]
\textbf{Hash Output} (Base64 strings, long fragments retained at both ends, middle lightly omitted):\\[4pt]
\par\centering
\begin{hashoutputbox}
\small\texttt{%
\textbf{FuzzyHash(A):} \codetext{64:lZk2s9Df0G7xR5tQ1bN3mP8vJ4hL$\cdots\cdots$6cX0zC2aS7dF9gH}\\[4pt]
\textbf{FuzzyHash(B):} \codetext{64:lZk2s9Df0G7xR5tQ1bN3mP8vJ4hL$\cdots\cdots$6cX0zD2aS7dF9gH}\\[4pt]
\textbf{FuzzyHash(C):} \codetext{64:qW7rT2yU5iO0pS3dF6gH9jK1lZ4xC$\cdots\cdots$8vB2nM5bN7}\\[4pt]
\textbf{Matching Statistics:} Compared with A, B differs by 1 character, while C differs by 28 characters.
}
\end{hashoutputbox}
\par
\end{expbox}

\begin{table}[htbp]
    \centering
    \small
    \renewcommand\arraystretch{1.15}
    \setlength{\tabcolsep}{2.4mm}
    \caption{Experimental indicators for the five hashing pipelines used in the case study.}
    \begin{tabular}{lcccc}
        \toprule
        \textbf{Algorithm Combination} & \textbf{A--B Similarity} & \textbf{A--C Similarity} & \textbf{MAP} & \textbf{Hash Time / Doc (s)} \\
        \midrule
        FlyHash + Jaccard & 0.778 & 0.000 & 1.00 & 0.0012 \\
        MinHash + Jaccard & 0.920 & 0.070 & 1.00 & 0.0028 \\
        SimHash + Manhattan & 0.333 & 0.031 & 1.00 & 0.0018 \\
        Winnowing + Jaccard & 0.714 & 0.000 & 1.00 & 0.0035 \\
        FuzzyHash + Levenshtein & 0.979 & 0.417 & 1.00 & 0.0022 \\
        \bottomrule
    \end{tabular}
    \label{Indicators-Summary}
\end{table}

%
%
%

\bibliographystyle{tmlr}
\bibliography{main}

\end{document}